\documentclass[sigconf, nonacm]{acmart} 

\usepackage{graphicx}
\usepackage{multirow}

\usepackage{devanagari}

\newcommand{\bheading}[1]{\hfill \break \noindent{\textbf{#1.}}}

\AtBeginDocument{%
  }

\setcopyright{acmlicensed}
\copyrightyear{2025}
\acmYear{2025}
\acmDOI{XXXXXXX.XXXXXXX}

\definecolor{darkgreen}{RGB}{60, 150, 50}
\definecolor{darkgrey}{RGB}{180, 180, 180}
\definecolor{darkblue}{RGB}{110, 130, 230}
\definecolor{bluishpurple}{RGB}{150, 85, 200}

\newcommand{\chiadd}[1]{\textcolor{black}{#1}}
\newcommand{\chihighlight}[1]{\textcolor{black}{#1}}
\newcommand{\chimohit}[1]{\textcolor{black}{#1}}
\newcommand{\chiaddagain}[1]{\textcolor{black}{#1}}
\newcommand{\chihighlightagain}[1]{\textcolor{black}{#1}}

\newcommand{\chidelete}[1]{}

\begin{document}

\title[ASHABot]{\ashabot{}: An LLM-Powered Chatbot to Support the Informational Needs of Community Health Workers}

\author{Pragnya Ramjee}
\authornote{Both authors contributed equally to this research.}
\orcid{0000-0003-0061-2624}
\affiliation{%
  \institution{Microsoft Research}
  \city{Bangalore}
  \country{India}}
\email{t-pramjee@microsoft.com}

\author{Mehak Chhokar}
\orcid{0009-0005-8778-9431}
\authornotemark[1]
\affiliation{%
  \institution{Khushi Baby}
  \city{Udaipur}
  \country{India}}
\email{mehak@khushibaby.org}

\author{Bhuvan Sachdeva}
\orcid{0009-0002-1946-684X}
\affiliation{%
  \institution{Microsoft Research}
  \city{Bangalore}
  \country{India}}
\email{b-bsachdeva@microsoft.com}

\author{Mahendra Meena}
\orcid{0009-0003-4649-2231}
\affiliation{%
 \institution{Khushi Baby}
 \city{Udaipur}
 \country{India}
 }
 \email{mahendra@khushibaby.org}

\author{Hamid Abdullah}
\orcid{0009-0004-4592-4060}
\affiliation{%
  \institution{Khushi Baby}
  \city{Udaipur}
  \country{India}}
   \email{hamid@khushibaby.org}

\author{Aditya Vashistha}
\orcid{0000-0001-5693-3326}
\affiliation{%
  \institution{Cornell University}
  \city{Ithaca}
  \country{USA}}
\email{adityav@cornell.edu}

\author{Ruchit Nagar}
\orcid{0000-0002-9461-8121}
\affiliation{%
  \institution{Khushi Baby}
  \city{Udaipur}
  \country{India}}
\email{ruchit@khushibaby.org}

\author{Mohit Jain}
\orcid{0000-0002-7106-164X}
\affiliation{%
  \institution{Microsoft Research}
  \city{Bangalore}
  \country{India}}
\email{mohja@microsoft.com}

\renewcommand{\shortauthors}{Ramjee et al.}

\newcommand{\ashabot}{\textit{ASHABot}}
\newcommand{\link}[1]{\textcolor{blue}{#1}}

\newcommand{\anonymousngo}{Khushi Baby}

\begin{abstract}
  Community health workers (CHWs) provide last-mile healthcare services but face challenges due to limited medical knowledge and training. This paper describes the design, deployment, and evaluation of \ashabot{}, an LLM-powered, experts-in-the-loop, WhatsApp-based chatbot to address the information needs of CHWs in India. Through interviews with CHWs and their supervisors and log analysis, we examine factors affecting their engagement with \ashabot{}, and \ashabot{}'s role in addressing CHWs' informational needs. We found that \ashabot{} provided a private channel for CHWs to ask rudimentary and sensitive questions they hesitated to ask supervisors. CHWs trusted the information they received on \ashabot{} and treated it as an authoritative resource. CHWs' supervisors expanded their knowledge by contributing answers to questions \ashabot{} failed to answer, but were concerned about demands on their workload and increased accountability. We emphasize positioning LLMs as supplemental fallible resources within the community healthcare ecosystem, instead of as replacements for supervisor support.

\end{abstract}

\ccsdesc[100]{Human-centered computing~Ubiquitous and mobile computing systems and tools}
\ccsdesc[500]{Applied computing~Health care information systems}
\ccsdesc[300]{Human-centered computing~Field studies}
\keywords{Chatbot, GPT-4, Experts-in-the-loop, Medical, Frontline Healthcare, HCI4D, ASHA, India}


\maketitle

\section{Introduction}

India's healthcare ecosystem relies on nearly a million Accredited Social Health Activist (ASHA) workers~\cite{asha2020-2021nhsrc}. 
ASHAs are volunteer community health workers who receive 3-4 weeks of initial training and are compensated through task-based incentives provided by the Government of India~\cite{chwcentral2024}.
They play a vital role in delivering essential health services, particularly in maternal and child health, within their communities~\cite{asha2020-2021nhsrc,chwcentral2024}.
However, ASHAs often encounter challenges in effectively delivering these services due to their limited medical knowledge~\cite{yadav2021gapsandneedsofchws}.
The reference material available to them are frequently insufficient for addressing complex queries of their care recipients~\cite{yadav2021gapsandneedsofchws}.
Additionally, ASHAs have limited opportunities for further training and upskilling~\cite{yadav2021gapsandneedsofchws}. 
For instance, although ASHAs attend monthly meetings at local hospitals where they can interact with doctors and participate in group training sessions~\cite{ismail2018solidarity,IPHS2022}, these sessions often lack adequate time to address ASHAs' individual queries comprehensively~\cite{ismail2018solidarity}.
Overall, there is a lack of skilled teachers, funds, space, and materials to effectively train and support ASHAs~\cite{yadav2017empoweringp2plearning,yadav2019leapscaffolding, Update_on_ASHA_Program_2013}.

ASHAs work under the supervision of Auxiliary Nurse-Midwives (ANMs), senior health workers who are formally employed and given 18 months of training by the government~\cite{kalne2022roleofchwsinindia,chwcentral2024,ismail2022caringfutures}. 
ASHAs often seek assistance from ANMs to perform their tasks and resolve doubts~\cite{yadav2017empoweringp2plearning,ismail2022caringfutures,yadav2019leapscaffolding}.
However, since a single ANM manages ASHAs across multiple villages, coordinating phone calls or in-person visits can be challenging~\cite{ismail2018solidarity}.
Furthermore, ANMs, with their limited formal education~\cite{lewin2010layhealthworkers,yadav2019leapscaffolding}, may not always have the necessary answers~\cite{javaid2017videolearninglhv}.
ASHAs also hesitate to discuss sensitive topics such as sexual health~\cite{ismail2018solidarity,rahman2021adolescentbot}.
As a result, many ASHAs' questions remain unanswered, and misunderstandings persist~\cite{yadav2019leapscaffolding}.

HCI researchers have investigated various technologies to address the informational needs of ASHA workers, including voice messaging~\cite{derenzi2017voiceandwebfeedback}, IVRS~\cite{yadav2017empoweringp2plearning}, feature phone apps~\cite{divya_mobileASHA_chi10}, and smartphone apps~\cite{yadav2019leapscaffolding,rodrigues2022communitycarehumaninloop,majhi2024smartphonerefreshers,panda2024echotelementoring}.
However, ASHAs often struggle with these tools due to a steep learning curve, largely stemming from their limited familiarity with digital technologies~\cite{ismail2018solidarity,ismail2019empowermentonmargins}.
Recently, chatbots have emerged as a more user-friendly alternative~\cite{yadav2019breastfeedingchatbot,karusala2023chatbasedinfoservice,mishra2023hindichatbot,pandey202qandasupport}.
However, despite their promise, the impact of previous chatbots has been constrained by their rule-based design, which limits natural language interaction, requires a steep learning curve, and often yields inaccurate responses.

In 2023, the release of LLM (Large Language Model)-powered chatbots marked a turning point, addressing these limitations and achieving higher adoption rates.
This trend has sparked renewed interest in using LLMs in healthcare~\cite{ramjee2024cataractbot,talk2care}.
However, applying LLMs to support healthcare in frontline settings remains largely unexplored and is still in early stages of design and evaluation~\cite{gangavarapu2024llmforhealthequity,mcpeak2024nigeriallmfordecisionmaking}.

In this work, we examine the potential of an LLM-powered chatbot to address the informational needs of ASHAs.
In collaboration with \anonymousngo{}, a non-governmental organization in Rajasthan, India, we designed and developed \ashabot{} using the open-source `Build Your Own expert Bot' framework~\cite{ramjee2024cataractbot}.
Key features of \ashabot{} include its integration with WhatsApp---a platform already familiar to ASHAs and ANMs~\cite{ismail2018solidarity}---and its ability to provide context-specific answers based on a custom knowledge base that incorporates government guidelines and ASHA handbooks~\cite{asha_handbooks}.
\ashabot{} values ANMs for their expertise, involving them as experts-in-the-loop to guide consensus-driven decisions on unfamiliar questions. 
The ANMs' answers are used to update the bot's knowledge base.
\ashabot{} supports Hindi and speech-based input and output, making it accessible to ASHAs and ANMs, who have varying literacy levels and digital skills~\cite{ismail2018solidarity,ismail2019empowermentonmargins}.
Given the application of LLMs in a high-stakes setting, we conducted iterative pilot testing to ensure the bot's usability and the accuracy of its responses.
We then conducted a \chidelete{in-the-wild }\chiadd{field}  deployment study with 20 ASHAs and 15 ANMs in the Salumbar district of Rajasthan, India. 
After two months of usage, we conducted interviews with ASHAs and ANMs to examine three key research questions: 
\textbf{(RQ1)} How did ASHAs and ANMs use \ashabot{}? 
\textbf{(RQ2)} What factors influenced their engagement with \ashabot{}? 
\textbf{(RQ3)} How effective is \ashabot{} in addressing ASHAs' information needs?

Our analysis of interviews and log data showed that ASHAs valued \ashabot{} for convenient, anytime access to detailed information, allowing them to learn at their own pace and save time.
The bot provided a private channel for ASHAs to ask rudimentary or sensitive questions that they might hesitate to ask ANMs or doctors.
ASHAs generally trusted the bot's answers, especially as \ashabot{} admitted when it did not know an answer.
Over time, \ashabot{} became an authoritative resource, influencing the decisions of both ASHAs and their care recipients.
Although most ANMs expanded their knowledge by providing answers to the bot, they expressed concerns about the added workload and being held accountable for their responses.
We conclude the paper by discussing design opportunities and challenges, such as personalization for ASHAs, incentivization for ANMs, and the importance of continuous training and support to address misconceptions in high-stakes frontline health settings. 
Our work makes the following contributions: 
\begin{itemize}
    \item We design, build, and deploy an LLM-powered, experts-in-the-loop chatbot on WhatsApp to support informational needs of low-skilled, low-literate ASHAs. 
    \item Through a field deployment and interviews with ASHAs and ANMs, we evaluate the efficacy of the bot and identify factors that shape ASHAs' and ANMs' engagement with it. 
    \item We discuss the limits and opportunities of using emergent AI technologies like LLMs to strengthen community health infrastructure. 
\end{itemize}
\section{Background and Related Work}
ASHAs play a crucial role in supporting the well-being of their communities, particularly in providing maternal and child health services, with each ASHA serving a population of $\sim$1,000 people~\cite{guidelinesforcommunityprocesses2014, whoun2008taskshifting}.
Their responsibilities include visiting pregnant women, assisting with checkups and deliveries, offering post-natal care, monitoring newborn health, and providing guidance on immunizations, family planning, and anemia prevention~\cite{yadav2019breastfeedingchatbot}.
While our research focuses on ASHAs in India, it is situated within the broader context of literature on technology supporting community health workers worldwide. 
Additionally, our work explores the use of AI in healthcare.
We summarize these topics below.

\subsection{Technology to Support Community Health Workers}
The HCI4D community has developed and deployed various solutions to improve the engagement of community health workers with care recipients.
These include tools for planning home visits~\cite{derenzi2012smstoimproveperformance,zurovac2011smsmalaria},
data collection~\cite{pal2017datadigitalbengal}, clinical decision-making~\cite{mcpeak2024nigeriallmfordecisionmaking},
health-related education through text messages~\cite{perrier2015smsforpregnantwomen,karusala2023chatbasedinfoservice,familyplanningchatbot_unfpa} and mobile videos~\cite{kumar2015videoeducationbyashas,molapo2017lesothovideolearning},
and feedback gathering~\cite{molapo2016participatorydesignvideolearning,okeke2019feedbackloopforcarerecipients}.
Researchers have also designed technologies to improve the performance of community health workers, 
for example by enabling supervisors to provide personalized feedback~\cite{derenzi2012smstoimproveperformance,whidden2018feedbackdashboard,karunasena2021datacollectiondiligence,Henry2016supervisionwhatsappkenya} and facilitating peer-comparison opportunities~\cite{derenzi2016feedbacklooptracking,derenzi2017voiceandwebfeedback}.
These technological interventions have not only helped towards legitimizing the roles of community health workers as healthcare providers---as they are often disregarded or unpaid~\cite{ismail2022caringfutures,kumar2015videoeducationbyashas,divya_mobileASHA_chi10}---but have also motivated them to improve their digital skills~\cite{ismail2019empowermentonmargins,medhi2012malnutritionchildrenmobilephones}.

In particular, the use of technology to address community health workers' information needs has been extensively studied~\cite{yadav2021gapsandneedsofchws}.
\chiadd{Prior research has explored Interactive Voice Response (IVR) systems for delivering automated refresher courses and quizzes~\cite{armman_mobile_academy,capacity_plus_senegal} and enabling ASHAs to participate in live audio training sessions with peers and trainers remotely~\cite{yadav2017empoweringp2plearning,divya_mobileASHA_chi10}.
However, IVR systems have significant limitations.
Sequential navigation requires users to listen to all options, leading to long wait times~\cite{nngroup2023phonetree}. 
Their menu structures are restrictive, particularly when navigating back to the root or previous menus~\cite{kim2012usabilityissuesinIVR}.
Limited support for local language speech technologies further forces reliance on Dual Tone Multi-Frequency (DTMF) inputs, which are unintuitive and slow~\cite{asthana2013IVRadaptiveinterfaces}. Additionally, the lack of persistent information reduces usability~\cite{nngroup2023phonetree}. 
These factors make IVR systems cost-prohibitive and inefficient~\cite{vashistha_voice_2023}.}
Visual solutions, which facilitate quick referencing, can overcome \chiadd{some of} these limitations.
Researchers have proposed various alternatives, including non-interactive videos~\cite{javaid2017videolearninglhv,vashistha_examining_2017}, smartphone-based games~\cite{majhi2024smartphonerefreshers}, custom platforms that leverage peer and supervisor knowledge for collaborative learning and telementoring~\cite{yadav2019leapscaffolding,rodrigues2022communitycarehumaninloop,panda2024echotelementoring}, and even virtual and augmented reality applications~\cite{majhi2023arplaytrainingashas, bhowmick2018pragati}.
Despite the innovative nature of these technologies, their adoption has been limited due to high costs, as well as the steep learning curve for community health workers, who have limited education and poor digital skills~\cite{ismail2018solidarity,ismail2019empowermentonmargins}.

Recently, chatbots have emerged as an effective tool to meet the information needs of community health workers~\cite{yadav2019breastfeedingchatbot,mishra2023hindichatbot,karusala2023chatbasedinfoservice}. 
With their natural language interface, 24x7 availability, support for multiple languages, and speech-based interactions, chatbots minimize the learning curve, making them accessible even to those with limited digital skills~\cite{farmchat_imwut18,ramjee2024cataractbot,bot-dis18}.
Chatbots can be easily deployed on familiar platforms such as WhatsApp~\cite{ismail2018solidarity,Henry2016supervisionwhatsappkenya}.
Before the advent of LLMs, researchers explored various rule-based chatbots.
For instance, \citet{yadav2019breastfeedingchatbot} used a participatory Wizard-of-Oz approach to examine the potential of chatbots for breastfeeding-related questions, highlighting that ASHAs saw it in the possible role of trainer to enhance their knowledge,
\citet{karusala2023chatbasedinfoservice} developed a WhatsApp-based service that allows nurses to manually respond to antenatal and postnatal care questions, and
\citet{mishra2023hindichatbot} created a rule-based chatbot using an FAQ to answer maternal and child health questions.
These non-LLM chatbots faced significant limitations~\cite{convey-chi18}, including (1) a limited scope that reduced their overall utility, and (2) difficulties in understanding natural language queries, leading to inaccurate responses.
In contrast, \ashabot{} uses the state-of-the-art LLM model, GPT-4, to provide instant and accurate responses to a wide range of queries from community health workers via WhatsApp, addressing these previous shortcomings. 
We now describe the scholarly work on designing, building, and evaluating AI technologies in community health context. 

\subsection{AI and Community Health}
Researchers have extensively explored the potential of AI to enhance the healthcare ecosystem~\cite{chen2018aiinmedicine,haleem2019aiinhealthcare,secinaro2021roleofaiinhealthcare}.
This research spans a variety of applications, including disease diagnosis~\cite{seismo, smartkc, respirenet}, clinical decision support~\cite{schaekermann2020aiformedicaldataassistant,zhang2020confidenceandtrustindecisionmaking}, improving communication between patients and healthcare providers~\cite{ramjee2024cataractbot,mittal2021chatbotfaqshospital,talk2care}, and optimizing hospital logistics~\cite{chakradhar2017optimaldrugsanddoses,bauerhenne2024robustappointmentschedulingwaiting,yildirim2024aiinicu}.
In the context of community healthcare, 
researchers have developed AI tools to help ASHAs and ANMs diagnose diseases~\cite{degreef2014bilicam,beede_human-centered_2020}, analyze rapid diagnostic tests~\cite{dell2014mobilehealthlowresource,park2020rapiddiagnosticsmartphonelowresource}, 
and manage patient-care~\cite{nishtala_missed_2020,rodrigues_digital_2022}. 
Studies have also examined ASHAs' knowledge of AI and their perception of AI-based diagnostic applications, and found that ASHAs possess low levels of AI understanding and a tendency towards overreliance on AI~\cite{okolo2021chwperceptionofAIinmhealth}. 
To ensure community health workers use AI technologies safely, researchers have explored methods to make AI more understandable to them~\cite{okolo2024explainableai} and designed novel tools, such as Explorable AI explanations, that allow low-skilled ASHAs to test their expectations of AI behavior against its actual behaviors, thereby reducing AI overreliance~\cite{solano-kamaiko_explorable_2024}. 

With the advent of LLMs, a growing body of work has explored how these emerging technologies can address the informational needs of community health workers~\cite{mcpeak2024nigeriallmfordecisionmaking,gangavarapu2024llmforhealthequity,antoniak2024nlpformaternalcare,perianez2024aiinhealth,talor2024malawilearningchws,ghadban2023llmsforcapacitybuildingofchws}.
For instance, \citet{ghadban2023llmsforcapacitybuildingofchws} developed an LLM that applies retrieval-augmented generation over a custom knowledge base, and \citet{gangavarapu2024llmforhealthequity} proposed a fine-tuned LLM.
Unlike our study, both these works focus on technical development and benchmarking rather than deployment with end-users.  
Similarly, \citet{talor2024malawilearningchws} developed a custom LLM-powered application to provide on-demand guidance to community health workers in Malawi. 
They evaluated the tool by conducting pilot focus group sessions, which indicated its potential for delivering point-of-care assistance in low-resource settings, but without a \chidelete{in-the-wild }\chiadd{field} deployment study.
\citet{mcpeak2024nigeriallmfordecisionmaking} explored the use of LLMs to aid Nigerian community health workers in decision-making by offering a `second opinion'.
This system relies solely on notes from patient interactions and does not address additional questions that may arise.
Our work contributes to this emerging body of research, by conducting a deployment study with 35 community health workers to assess the factors influencing their interaction with \ashabot{}, an LLM-powered, WhatsApp-based expert-in-the-loop chatbot designed to meet their information needs.

Finally, our \chidelete{in-the-wild }\chiadd{field} study was informed by design guidelines for AI in frontline health~\cite{okolo2021chwperceptionofAIinmhealth,ismail2021aiinglobalhealth,okolo2024explainableai}.
For example, \citet{ismail2021aiinglobalhealth} advocate for strong partnerships with community organizations and \citet{okolo2024explainableai} emphasize the need for training and ongoing technical support during deployment.
Throughout the process, we collaborated closely with our partner organization \anonymousngo{}, who played a crucial role in onboarding community health workers onto \ashabot{} and providing them with continuous support. 
In the following sections, we describe the design of \ashabot{} and the methodology used to evaluate its efficacy in addressing the informational needs of ASHAs.

\section{System Design}
\label{systemdesign}

\begin{figure*}
  \includegraphics[width=\textwidth]{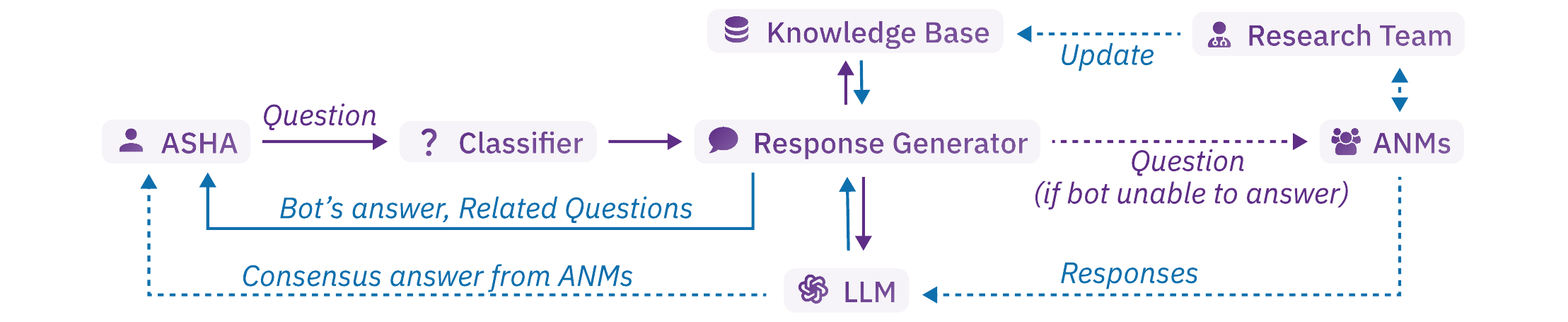}
  \caption{Flow diagram of the \ashabot{} system.}
  \Description{Flow diagram of the ASHABot system. When an ASHA asks a question, the bot responds by searching a doctor-curated knowledge base. It passes the retrieved information to GPT-4, which generates an answer as well as three Related Questions for the ASHA. If the bot cannot find an answer within its knowledge base, it forwards the question to multiple ANMs, and creates a consensus answer from their responses which it then sends back to the ASHA. The knowledge base is updated by our research team, based on the ANMs' inputs.}
  \label{fig:ashabotflow}
\end{figure*}

\ashabot{} is an LLM-powered, expert-in-the-loop, WhatsApp-based chatbot system.
It is built on an open-sourced framework called BYOeB\footnote{Build Your Own expert Bot: \href{https://github.com/microsoft/byoeb}{\link{https://github.com/microsoft/byoeb}}}, which was originally designed, developed, and deployed in a hospital setting for cataract surgery patients by \citet{ramjee2024cataractbot}.
\ashabot{} (Figure \ref{fig:ashabotflow}) responds to ASHAs' questions by searching a doctor-curated knowledge base and using GPT-4 to generate an answer from the retrieved information.
If no answer is found, it forwards the question to ANMs, gathers their responses, generates a consensus answer, and sends that back to the ASHA.
Additions to the knowledge base are based on the ANMs' answers and approved by medical doctors on our research team, enabling the bot to improve over time.

\begin{figure*}
  \includegraphics[width=\textwidth]{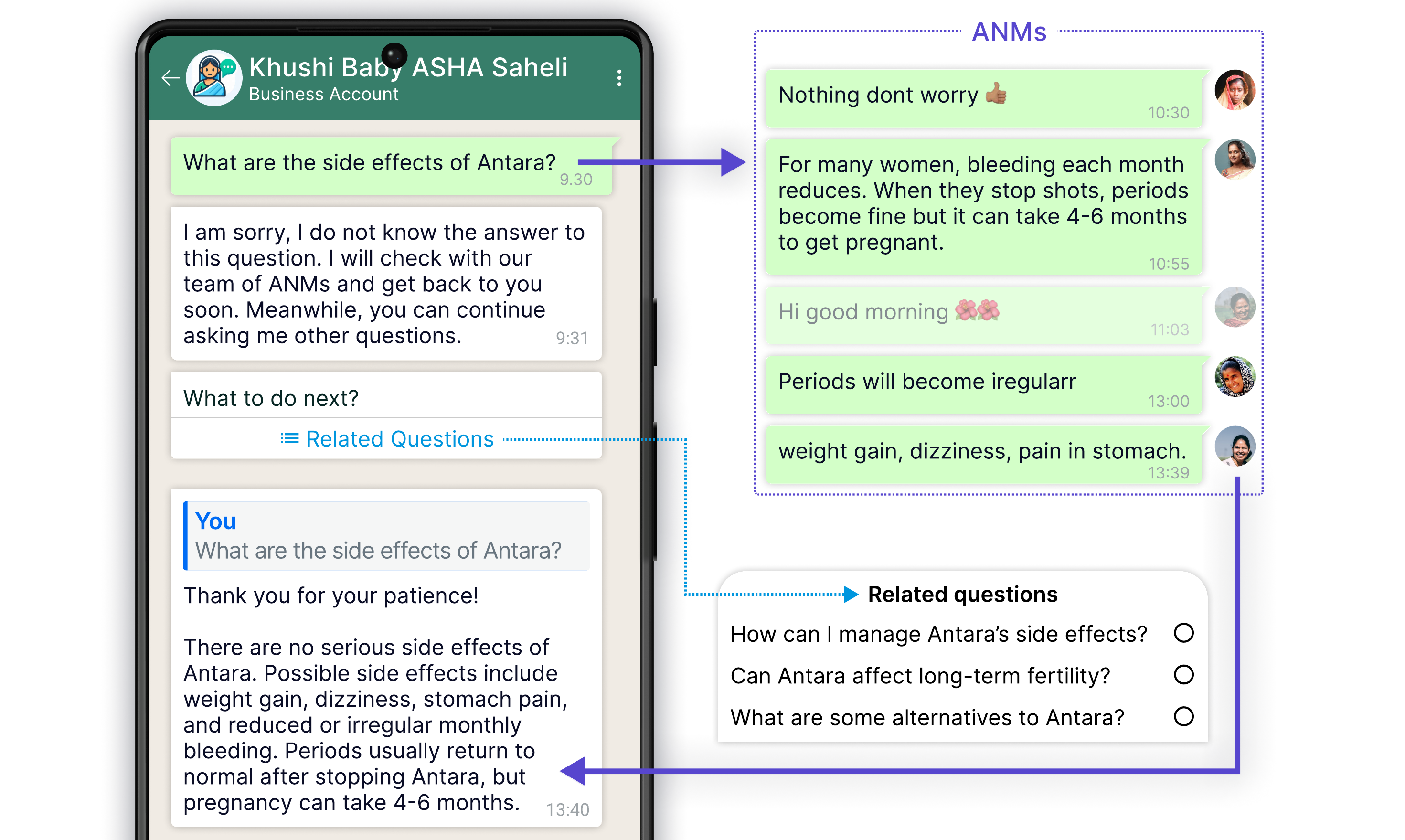}
  \caption{When \ashabot{} cannot answer a question using its knowledge base, it sends that question to multiple ANMs. It identifies the relevant information from their responses and generates a consensus answer, which it sends back to the ASHA. While we present this example in English, we note that all ASHAs and ANMs interacted with \ashabot{} in Hindi.}
  \Description{ASHABot on WhatsApp, showing an ASHA's question on the side effects of the Antara injectable contraceptive. ASHABot is unable to generate an answer from its knowledge base, and informs the ASHA that it will check with the team of ANMs and revert. It also provides the ASHA with 3 Related Questions which can be asked in the meantime: 'How can I manage Antara's side effects?', 'Can Antara affect long-term fertility?', and 'What are some alternatives to Antara?'. The bot sends the ASHAs' question to multiple ANMs, and the figure shows 5 ANMs' responses: 'Nothing dont worry', 'For many women, bleeding each month reduces. When they stop shots, periods become fine but it can take 4-6 months to get pregnant.', 'Hi good morning', 'Periods will become iregularr', and 'weight gain, dizziness, pain in stomach.'. ASHABot identifies relevant information from these responses, such as by discarding the morning greeting, and creates a consensus answer, which it sends back to the ASHA: 'There are no serious side effects of Antara. Possible side effects include weight gain, dizziness, stomach pain, and reduced or irregular monthly bleeding. Periods usually return to normal after stopping Antara, but pregnancy can take 4-6 months'.}
  \label{fig:ashabotsystem}
\end{figure*}

The \ashabot{} system incorporates a few key modifications from the original BYOeB framework.
First, in BYOeB, each query is sent to a single expert for verification, which is well-suited to hospital settings where doctors have extensive medical training and expertise.
In the \ashabot{} context, ANMs serve as the experts.
They typically have limited formal education, supplemented by only 18 months of training without a professional degree~\cite{lewin2010layhealthworkers, yadav2019leapscaffolding}.
To ensure a more reliable and comprehensive verification process, \ashabot{} uses a crowdsourcing approach for answers that cannot be found in its knowledge base, aggregating responses from multiple ANMs (Figure \ref{fig:ashabotsystem}).
The final answer shared with the ASHA is determined by adopting the most common response among the ANMs.
In cases of conflicting opinions, majority voting is employed.
If fewer than three relevant responses are received, or if a tie occurs during majority voting, the system automatically forwards the query to additional ANMs.
The detailed prompt used in this consensus-obtaining process is available in Appendix \ref{appendix:prompts:consensus}.
This prompt was refined through multiple iterations over a synthetic dataset that includes 
questions posed by ASHAs and answers provided by ANMs.

Second, in the BYOeB framework, every LLM-generated response is sent to an expert for verification. 
\chiadd{In \ashabot{}, however, only questions that result in an `\textsf{I don't know}' response---indicating the chatbot cannot generate an answer from the curated knowledge base---are forwarded to ANM experts (Figure \ref{fig:ashabotsystem}).
This design choice was made to balance scalability with the demanding schedules of ANMs, who are often busy and overstretched~\cite{yadav2017empoweringp2plearning}. To further reduce `\textsf{I don't know}' responses and minimize the workload of ANMs, we iteratively improved the custom knowledge base.}
Initially, the knowledge base was prepared using the seven-part ASHA training module~\cite{asha_handbooks}. 
\anonymousngo{} then collected 197 questions frequently asked by ASHAs and generated responses from \ashabot{} for each question.
\ashabot{} was unable to answer 76 of these questions, primarily due to gaps in the knowledge base. 
To address this, we supplemented \ashabot{}'s knowledge base with additional documents specific to those topics and iterated the process twice.

Third, for questions asked by ASHAs using the speech modality, \ashabot{} provides ANMs with both the original audio recording and the transcribed text version.
This approach addresses limitations identified by \citet{ramjee2024cataractbot} where only the text version was shared with experts, leading to a few instances of misunderstandings due to transcription errors.
To accommodate varying levels of (technological) literacy, ANMs can provide their answers via either text or voice messages, unlike the text-only approach used for experts by \citet{ramjee2024cataractbot}.

Lastly, we intentionally defined a female gender for \ashabot{}.
In deference to Indian cultural sensitivities about conversations related to pregnancy and childcare~\cite{varkey2004meninmaternitycare}, we avoided an explicit male presence in our system.
As shown in Figure \ref{fig:ashabotsystem}, \ashabot{}'s display name on WhatsApp was `\textit{\anonymousngo{} ASHA Saheli}', \chiadd{
which translates to `\textit{a female friend of ASHAs}' in English. }

\subsection{Pilot Deployment and \chiadd{Response Evaluation}}
\label{systemdesign:pilot}
\chihighlightagain{In March-May 2024, we deployed \ashabot{} with test users from \anonymousngo{} and our research team with 19 onboarded as ASHAs and 10 as ANMs.
Test users from \anonymousngo{} included field workers, who serve as direct liaisons with ASHAs and ANMs.}
\chidelete{To gather feedback, we conducted interviews with two male field workers (average age: 29.0±4.24 years). 
Both participants were fluent in Hindi and Mewari, and the interviews were conducted in Hindi. 
The first author transcribed and translated these interviews into English.}
\chihighlight{\chihighlightagain{These field workers asked 374 questions, reviewed \ashabot{}'s responses, and provided feedback on their accuracy and completion.}
\chihighlightagain{Of these, 273 responses were marked as accurate and complete, \chidelete{13 }17 as accurate but incomplete, \chidelete{13 }9 as inaccurate, and 75 were `\textsf{I don't know}' responses.
Of the inaccurate responses, 3 were due to discrepancies (e.g., dosage inconsistencies) in the government-approved documents.
We addressed these by manually updating the knowledge base with doctor-verified answers, and instructing the bot to prioritize recent updates while generating responses.}}
The\chidelete{se inaccurate or} `\textsf{I don't know}' responses were mainly due to translation and transcription errors.
For example, 
\ashabot{} mistranslated `\textsf{ASHA}' in the Hindi question to `\textsf{hope}' in English, resulting in an `\textsf{I don't know}' response. 
\chihighlight{To resolve this issue, we experimented with different query inputs to the GPT-4 model: (a) the original Hindi query, (b) the English-translated query, and (c) both the English-translated query and the Hindi source query. 
The number of `\textsf{I don't know}' responses for these methods were 32, 35, and 32, respectively.
We decided to input both Hindi and English queries into the system. 
This approach further distinguishes our method from that of \citet{ramjee2024cataractbot}, who only used English queries.} 

\chihighlight{To address transcription failures, we developed a dictionary, initially containing 4 word pairings. 
For instance, the term `Antara', referring to an injectable contraceptive (medically known as Depot Medroxy Progesterone Acetate~\cite{ray2024antara}), was often misheard by the bot as `\textsf{Anna}'.
Second, to assist ASHAs when their audio query resulted in an `\textsf{I don't know}' response, we included a transcription of the audio processed by \ashabot{} alongside the `\textsf{I don't know}' message. 
This allowed ASHAs to understand how their question was interpreted and to refine their question accordingly.}
Pilot feedback also indicated that the Hindi used by the bot was perceived as ``\textit{too pure}''.
As a result, instead of using Azure for English-to-Hindi translation, we experimented using GPT-4 for translation.
We requested \anonymousngo{} to evaluate the understandability of 26 bot responses translated by both Azure and GPT, using a 5-point Likert scale, where 1 represented `Very easy' and 5 represented `Very hard'. 
The average rating for Azure was 1.92$\pm$1.29, while for GPT, it was 1.58$\pm$0.90. 
Hence, we decided to use GPT for translation of the output.

\subsection{Evaluation of \ashabot{}'s \chiadd{Responses by Medical Professionals}}

Given the deployment of \ashabot{} in a high-stakes healthcare context, 
we rigorously evaluated its responses using multiple approaches (Table \ref{tab:eval-bot}).
\begin{table*}
\centering
\caption{\textbf{Evaluation of \ashabot{}'s answers.} First, we used an LLM prompt with six metrics to compare the bot's answers to Ground Truth (GT) answers independently generated by three human experts (Doctor, CHO, and ANM). Then, two doctors manually evaluated the bot's answers as per their knowledge, using three metrics.}
\label{tab:eval-bot}
\resizebox{\textwidth}{!}{%
\begin{tabular}{l|c|c|c|c|c|c} 
\hline
 &
  \textbf{Accuracy} &
  \textbf{Subset} &
  \textbf{Completeness} &
  \textbf{Conciseness} &
  \textbf{Clarity} &
  \textbf{Structure} \\ \hline
\textbf{GT = Doctor} & 87.88\%$\pm$26.32\% & 72.67\%$\pm$16.26\% & 67.33\%$\pm$10.74\% & 100.00\%$\pm$0.00\% & 98.00\%$\pm$8.00\% & 74.00\%$\pm$24.09\% \\
\textbf{GT = CHO} &
  93.65\%$\pm$20.05\% &
  65.99\%$\pm$22.27\% &
  63.27\%$\pm$19.74\% &
  100.00\%$\pm$0.00\% &
  98.67\%$\pm$6.60\% &
  65.97\%$\pm$29.89\% \\
\textbf{GT = ANM} &
  96.08\%$\pm$16.17\% &
  63.95\%$\pm$22.18\% &
  59.86\%$\pm$22.45\% &
  99.33\%$\pm$4.76\% &
  96.00\%$\pm$11.04\% &
  61.90\%$\pm$30.46\% \\ \hline
\textbf{Doctor 1} &
  93.65\%$\pm$19.73\% &
  96.30\%$\pm$10.56\% &
  89.42\%$\pm$15.64\% &
   &
   &
   \\
\textbf{Doctor 2} &
  85.00\%$\pm$19.75\% &
  82.00\%$\pm$17.36\% &
  94.67\%$\pm$12.28\% &
   &
   &
   \\ \hline
\end{tabular}%
}
\end{table*}

First, we randomly selected 50 questions posed by test users and compared \ashabot{}'s responses against responses independently generated by three expert groups: doctors, ANMs, and Community Health Officers (CHOs)\footnote{CHOs are frontline health workers with a bachelor's degree in nursing, who supervise ANMs in rural areas~\cite{nhsrc2021cho}.}. 
These expert-generated responses served as the ground truth for our analysis.
For this evaluation, we used a prompt (Appendix \ref{appendix:prompts:evaluation}) which included six metrics based on those identified in prior literature~\cite{abd2020metricsforhealthcarechatbots}.
The metrics were (1) `Accuracy', the correctness of the information in the answer, compared to the ground truth, (2) `Subset', the degree to which the information in the answer is contained within the ground truth), (3) `Completeness', the degree to which the information in the ground truth is contained within the answer, (4) `Conciseness', the degree of repetition in the answer, (5) `Clarity', the simplicity of medical information in the answer, and (6) `Structure', the logical arrangement of information in the answer compared to the ground truth.
Each metric used a 3-point scale, except for `Accuracy', which used a 2-point scale but included an `NA' category for indeterminate answers which included information not present in the ground truth.
We refined the evaluation prompt iteratively by using it to evaluate \ashabot{}'s answers to randomly chosen questions.
Each iteration involved comparing the LLM's evaluation with a manual evaluation, conducted by the first author \chiadd{using the same six metrics.}

\chihighlight{
In addition to the automated evaluation, two doctors also manually assessed the responses generated by \ashabot{} based on three metrics: `Accuracy', `Subset, and `Completeness'. 
Doctor 1 evaluated 63 responses, and Doctor 2 reviewed 100.
The results of these evaluations are also summarized in Table \ref{tab:eval-bot}. 
Our evaluations found that \ashabot{} achieved an accuracy of 85\% or higher in all cases, which provided us  with the confidence to deploy \ashabot{} among ASHAs and ANMs in the field, which we describe next.
}
\section{Field Deployment and Evaluation}
With the help of \anonymousngo{}, we deployed \ashabot{} with ASHAs and ANMs, and conducted an IRB-approved mixed-methods user study to understand their engagement with \ashabot{}.
Below, we detail our deployment approach and evaluation methodology, including participant recruitment, demographic information, interviews, and data analysis methods.
 
\subsection{Field Deployment}
We deployed \ashabot{} between May-August 2024 to ASHAs and ANMs within the Salumbar block of Salumbar district in Rajasthan, India's largest state. 
The Salumbar district was selected due to \anonymousngo{}'s established presence there, allowing them to obtain the necessary government permissions for deployment and
identify a pool of potential study participants. 
\chiadd{In Salumbar, ASHAs usually work for 6 hours daily, visiting $\sim$10 households in their village.
On the other hand, ANMs work in local clinics (subcenters), with 1-2 ANMs providing basic healthcare and family planning services to 4-5 villages.
ASHAs typically meet their ANM supervisor once a week at the subcenter.
At the end of each month, members from all subcenters within a sector--usually 25-30 ASHAs, 5-7 ANMs, and 2-3 CHOs--gather at the primary health center for a review meeting led by a doctor. 
Each sector also maintains an official WhatsApp group, primarily for announcements, with ASHAs typically contributing field photos as proof of their work or raising issues like delayed payments.
Beyond this, ASHAs actively participate in informal WhatsApp groups for emotional support, 
where they discuss common challenges such as limited incentives or overwhelming workloads.}

\chihighlight{A field worker from \anonymousngo{}, who had prior experience working with ASHAs and ANMs, facilitated participant recruitment.
The field worker introduced the participants to \ashabot{} and explained the study protocol, which involved using \ashabot{} for three months and participating in an interview with our research team during that period.
A total of 20 ASHAs and 15 ANMs agreed to participate in the study.
Their phone numbers were added to \ashabot{}'s database, and the bot sent them `welcome messages' with a `\textsf{Yes}' option as consent for opt-in.}
\chiadd{The field worker then conducted brief training sessions ($\sim$15 minutes) with groups of 2-5 participants. For the ASHAs, the training included an overview of the bot's purpose, demonstrations of how to ask questions using text, voice, and related-question inputs, and a reminder that the bot's responses might occasionally be inaccurate. ASHAs were encouraged to report inaccuracies and consult their ANMs if unsure. 
For the ANMs, the training highlighted their role in addressing ASHAs' questions the bot could not answer. The field worker demonstrated input methods and explained how their responses would be relayed anonymously to ASHAs after reaching a group consensus.}
Throughout the 3.5-month deployment, the field worker maintained regular phone contact with all participants, addressing any issues or questions and offering encouragement. 
One of the authors also visited five participants in-person during the first two weeks of deployment to gather early feedback and address any questions or concerns.

\chiadd{
We logged participants interactions with \ashabot{} for research analysis.
To promote open communication, ASHAs' conversations with \ashabot{} were not shared with their supervisors, ensuring privacy and avoiding additional burdens on supervisors to review these logs.
Participants were not compensated for this study, although they received financial support, including covering their mobile data cost, from \anonymousngo{} as part of a separate project deployment.}

\subsection{Semi-structured Interviews}

After ASHAs and ANMs had used the bot for two months, we interviewed them over a one-week period.
Following \anonymousngo{}'s recommendation, we focused on active \ashabot{} users. 
Specifically, we included ASHAs who had asked at least 10 questions and ANMs who had answered at least 2 questions. 
As a result, we interviewed 18 out of the 20 ASHAs and 10 out of the 15 ANMs.


\begin{table*}
\centering
\caption{Demographic details of interviewed ASHAs and ANMs}
\label{tab:demo-participants}
\begin{tabular}{r|ll}
\hline
                                & \textbf{ASHAs (n = 18)} & \textbf{ANMs (n = 10)} \\ \hline
\textbf{Gender}                 & Female                  & Female                 \\
\textbf{Age (yrs)}              & 34.7$\pm$6.21           & 33.3$\pm$8.69          \\
\textbf{Education} &
  \begin{tabular}[c]{@{}l@{}}\textless{}10th grade: 3, 12th grade: 9, \\ Bachelors: 4, Masters: 2\end{tabular} &
  \begin{tabular}[c]{@{}l@{}}\textless{}10th grade: 1, 12th grade: 4, \\ Bachelors: 2, Masters: 3\end{tabular} \\
\textbf{Work experience (yrs)}  & 10.7$\pm$4.81           & 7.90$\pm$9.13          \\
\textbf{Smartphone usage (yrs)} & 5.75$\pm$3.03           & 5.15$\pm$3.49          \\
\textbf{Languages known} &
  \begin{tabular}[c]{@{}l@{}}Hindi: 18, Mewari: 16, Wagdi: 1, \\ Dawadi: 1, Gujarati: 1, Sindhi: 1\end{tabular} &
  \begin{tabular}[c]{@{}l@{}}Hindi: 10, Mewari: 9, Wagdi: 2, \\ Marwari: 1, Gujarati: 1, English: 1\end{tabular} \\ \hline
\end{tabular}
\end{table*} 

The first two authors conducted semi-structured interviews in Hindi.
One researcher led the interviews, while the other took extensive notes.
To ensure participants' convenience, the interviews were held in-person at local nursery schools, health centers, hospitals, or participants' homes.
The interviews began with chit-chat to create a welcoming environment and then explored the participants' overall experience with \ashabot{}.
We focused on specific instances of why, when, and how they used the bot, encouraging comparisons with their existing methods for seeking or disseminating information.
Discussions included specific features they liked and disliked, suggestions for improvement, and their views on \ashabot{}'s trustworthiness and accuracy.
At the end of the interview, any misconceptions about the chatbot were clarified.
Each interview lasted $\sim$30 minutes.
With participants' consent, we audio-recorded 25 interviews; three participants did not consent to recording.
After the interviews, participants continued using \ashabot{} for an additional month.
Table \ref{tab:demo-participants} presents the demographic details of the interviewed participants.

\subsection{Data Analysis}
The first two authors translated and transcribed the 25 interview recordings, totaling 12 hours and 10 minutes, into English.
They also digitized their respective field notes.
The first author then conducted a thematic analysis of these transcripts and field notes, applying open-coding on a line-by-line basis.
Throughout the analysis, four authors regularly met to review, merge, or eliminate codes. 
We identified 124 unique codes in the ASHA transcripts and 96 in the ANM transcripts.
Through peer debriefing, we organized our findings around seven key themes, such as `Learning', `Authoritativeness', and `Motivators for Usage'.
We supplemented these qualitative insights with quantitative data derived from the interaction logs, including the count, type, and frequency of ASHAs' questions, as well as responses from \ashabot{} and ANMs.

\subsection{Ethics}
\chiadd{
Community health care presents a sensitive landscape within which, without proper safeguards, research activities could lead to unintended negative consequences. 
We took a number of steps to design, deploy, and evaluate \ashabot{} ethically and responsibly. 
To begin with, our team included HCI researchers, medical doctors, and experienced staff from \anonymousngo{}, with over a decade of experience working directly with ASHAs and ANMs. 
This interdisciplinary expertise, coupled with a commitment to community-engaged research, guided the design and evaluation of \ashabot{} to meet the specific needs of ASHAs and ANMs.}

\chiadd{
Recognizing the potential for LLMs to hallucinate~\cite{gould2024checkllm}, we conducted iterative pilot testing to refine the bot's knowledge base and periodically evaluated its performance.
Collaborating with medical doctors and field workers, we incorporated their feedback to ensure that the bot's responses were medically accurate, complete, and contextually and linguistically appropriate. 
We also systematically evaluated \ashabot{}'s performance via manual reviews of its responses by field staff and two doctors on our team, and also by comparing its answers to those provided independently by doctors, ANMs, and CHOs. 
It was only after receiving quantitative reassurance of the bot's accuracy as well as approval from the partner organization and local government officials that we deployed \ashabot{} to ASHAs and ANMs in the field.
To further enhance reliability, we employed conservative confidence thresholds for \ashabot{}'s responses. Questions beyond the bot's knowledge were forwarded to ANMs, whose collective responses were synthesized and relayed to ASHAs. 
Considering prior evidence of high AI overreliance among ASHAs~\cite{okolo2021chwperceptionofAIinmhealth,okolo2024explainableai}, participants underwent training before deployment. 
During these sessions, ASHAs were cautioned about potential inaccuracies in the bot’s responses and advised to consult their supervisors when in doubt.}

\chiadd{
Finally, throughout the deployment, our research team and \anonymousngo{} staff actively monitored the questions posed by ASHAs and the responses provided by both \ashabot{} and ANMs. 
This continuous oversight ensured the system's effectiveness and maintained its alignment with the needs of the community.
}

\subsection{Positionality}
All the authors are from India, and four have extensive experience conducting fieldwork in disadvantaged Indian communities.
Three authors have been part of the partner organization, \anonymousngo{}, for several years, and one is a field worker who lives and works closely with community health workers \chiadd{on a daily basis.} 
We recognize, however, that our socioeconomic status creates a power imbalance between us and our participants, who are women working in low-income,  highly patriarchal contexts.
A male field worker \chiadd{with strong community connections} attended the beginning of the interviews to build rapport but then stepped away to allow participants to discuss sensitive topics freely. 
\chihighlight{To create a comfortable and respectful environment for ASHAs and ANMs, two female researchers conducted the interviews in Hindi, addressing participants with respectful terms like ``\textit{ji}'', and ``\textit{ma'am}'', mirroring the participants' own use of these terms. 
This approach elicited nuanced responses.
In this work, we aim to use our relative privilege to amplify the voices of our participants, highlight their information needs, and critically examine the potential of emerging LLM-powered technologies to strengthen the community healthcare ecosystem.}
\section{Findings}
\label{findings}

Our analysis revealed that \ashabot{} effectively met the information needs of ASHAs and served as an authoritative resource during their fieldwork. 
Section \ref{findings:rq1} presents usage statistics extracted from the interaction logs to describe how ASHAs and ANMs engaged with the bot (RQ1).  
Section \ref{findings:rq2} describes key factors like convenience, human infrastructure limitations, privacy, and accountability that affected ASHAs' and ANMs' engagement with the bot (RQ1 and RQ2).
Finally, Section \ref{findings:rq3} describes how \ashabot{} supported ASHAs' learning and examines issues of trust, expertise, and authority (RQ2 and RQ3).

\subsection{Bot Usage by ASHAs and ANMs}
\label{findings:rq1}

\begin{table*}[]
\centering
\caption{Question Topics and Examples}
\label{tab:questiontopics}
\resizebox{\textwidth}{!}{%
\begin{tabular}{l|p{0.45\textwidth}|p{0.45\textwidth}}
\hline
\textbf{Question Topic} &
  \textbf{Example (Original Question)} &
  \textbf{Example (English Translation)} \\ \hline
\multirow{2}{*}{Pregnancy} &
  `{\dn agr ek gB\0vtF mEhlA ecAe\?ivF pA\<E)EVv ho to usk\? Ele \3C8wA kr\?{\qva}{\rs ?\re}}' &
  `\textsf{What to do if a pregnant woman is HIV-positive?}' \\  
 &
  `{\dn gB\0vtF k\? EktnF bFpF km h\4 to mAnA jAtA.}' &
  `\textsf{For a pregnant woman, what BP level is considered low?}' \\ \hline
\multirow{2}{*}{Childcare} &
  `{\dn is b\3CEw\? ko j\306wm k\? \7{t}r\2t pFElyA hotA h\4 vo k\4s\? ho jAtA h\4{\rs ?\re}}' &
  `\textsf{This child has jaundice immediately after birth, how does it happen?}' \\  
 &
  `{\dn ek b\3CEwA do Eklo kA h\4 to uskF d\?KBAl Eks trh krnF cAEhe{\rs ?\re} Gr{\rs ?\re}}' &
  `\textsf{If a child weighs two kilos, how should he be cared for? At home?}' \\ \hline
Family planning &
  `{\dn a\2trA i\2j\??fn lgvAn\? k\? bAd mEhlA kF mhAvArF \8{\3C8w}\2 zk jAtF h\4}' &
  `\textsf{Why does a woman's period stop after getting Antara injection?}' \\ \hline
\multirow{2}{*}{Societal issues} &
  `{\dn E\3FEwy\2kA aOr E\3FEwy\2kA k\? pEt kA JgwA ho jAtA h\4. \3C8wA krnA pwtA h\4 E\3FEwy\2kA{\rs ?\re}}' &
  `\textsf{Priyanka and Priyanka's husband get into a fight. What does Priyanka have to do?}' \\  
 &
  `{\dn bAl EvvAh honA cAEhe yA nhF{\qva}{\rs ?\re}}' &
  `\textsf{Should there be child marriage or not?}' \\ \hline
\multirow{2}{*}{General healthcare} &
  `{\dn eEsEXVF hotF h\4 to \3C8wA krnA pwtA h\4{\rs ?\re}}' &
  `\textsf{What to do if acidity occurs?}' \\  
 &
  `\textsf{Kya aayushman yojna pure bharat me lagu hai}' &
  `\textsf{Is Ayushman Yojana applicable to the whole of India?}' \\ \hline
General knowledge &
  `{\dn sAibr mfFn \3C8wA h\4{\rs ?\re}}' &
  `\textsf{What is a cyber machine?}' \\ \hline
\end{tabular}%
}
\end{table*}

\begin{figure*}[]
  \includegraphics[width=0.8\textwidth]{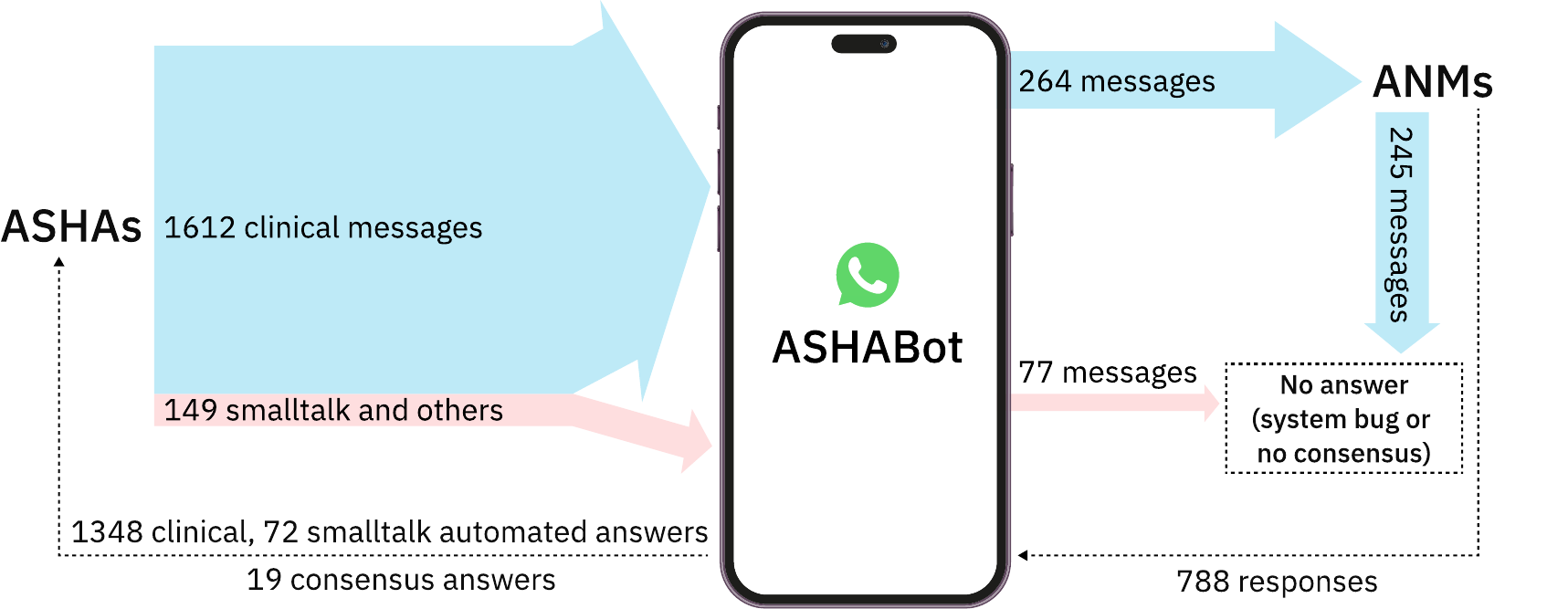}
  \caption{\chiadd{\ashabot{} deployment statistics.}}
  \Description{Flow diagram of key deployment statistics, with arrow widths proportional to number of messages. ASHAs sent 1,612 clinical messages and 149 smalltalk and other messages. The bot automatically responded to 1348 of the clinical messages and 72 of the smalltalk messages, but 77 messages were ignored due to a system bug. 264 clinical messages were forwarded to ANMs as the bot could not generate an answer from its knowledge base. ANMs sent a total of 788 responses, which were combined to form 19 consensus answers sent to ASHAs. The remaining 245 clinical messages sent to ANMs did not receive enough responses to generate consensus answers.}
  \label{fig:ashabotdeploymentresults}
\end{figure*}

Over the 107 days of deployment, ASHAs and ANMs used \ashabot{} extensively \chiadd{(Figure \ref{fig:ashabotdeploymentresults})}. 
ASHAs sent 1,761 messages, averaging 88.05± 123.01 messages and 0.83±1.16 daily messages, per ASHA.
The high standard deviations are due to two ASHAs who sent no messages, six who sent 100+, and one (ASHA1) who sent 555 messages. 
ASHAs sent 37.0\% of messages between 5-9 AM, using the early hours to prepare for their workday or reflect on the previous one.
ASHAs prioritized ease of use in their interactions.
They sent 1,040 messages by tapping on the `Related Questions' that the bot provided them with, recorded 470 audio messages, and typed just 251 text messages.
\chihighlight{Table \ref{tab:questiontopics} shows question topics and corresponding examples of questions asked by the ASHAs.
\ashabot{} classified 1,612 (91.5\%) of ASHAs' messages as clinical queries,} covering a broad range of topics beyond pregnancy and childcare. 
Of the remaining messages, 72 were smalltalk, such as `\textsf{Hii}', `\textsf{Thank you} \includegraphics[height=0.65\baselineskip]{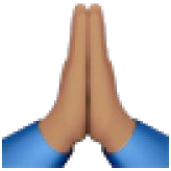}', and `{\dn aAps\? \3FEw\397w \8{p}Ckr b\7{h}t aQCA}' [`It's great to ask you questions'] (ASHA4), and 77 messages were ignored due to a system bug \chiadd{(Figure \ref{fig:ashabotdeploymentresults})}.
\chihighlight{\ashabot{} responded with `\textsf{I don't know}' to 264 clinical questions, forwarding them to ANMs.}

When \ashabot{} prompted ANMs to answer a question, 
they frequently ignored that.
Out of the 6915 prompts sent (461±4.40 prompts/ANM), ANMs responded `\textsf{Yes}' to only 1491, therefore receiving 99.4±99.78 questions/ANM.
In response to the questions, ANMs sent 788 messages overall (52.53±68.93 responses/ANM, and 0.46±0.63 daily responses/ANM).
Similar to ASHAs, three ANMs sent no messages, three sent 100+, while one (ANM1) sent 244. 
ANMs sent 37.9\% of their responses before their work day began at 9AM, with activity peaking at 1PM during lunch breaks.
After 5PM, ANMs did not send any responses.
Unlike ASHAs, only the minority of ANM responses used audio modality rather than text (3.2\%, or 1.67±3.06 responses/ANM).

To generate a consensus answer, \ashabot{} required agreement from at least three ANMs.
As ANMs averaged 2.28±1.76 responses per question, consensus was reached for just 19 questions.
Individual ANMs took 30.75±66.46 hours to respond, and the consensus answer was shared with the ASHA within 58.59±99.18 hours.
By that time, ASHAs typically had already found answers elsewhere and often overlooked \ashabot{}'s consensus answers.
This suggests that ASHAs primarily relied on \chihighlight{\ashabot{}'s automated answers, which accounted for 83.6\% of responses to clinical questions.}
A poll was sent along with 20\% of the bot's answers, asking `\textsf{Is this answer helpful?}'.
95.7\% of ASHAs' responses were `\textsf{Yes}', suggesting its effectiveness.
To more deeply understand the ASHAs' and ANMs' lived experiences with \ashabot{}, we proceed to examine factors that shaped their engagement with it.

\subsection{Factors Impacting Users' Engagement with the Bot}
\label{findings:rq2}

\subsubsection{Motivators for Bot Usage}
\label{findings:rq2:motivators}

ASHAs described several motivations to use \ashabot{}. 

\bheading{Convenience}
ASHAs found the bot highly convenient, allowing them to ask questions anytime, anywhere, without waiting for or traveling to meetings with ANMs and doctors.
Previously, ASHAs might forget their questions that they intended to ask their ANMs later, 
but \ashabot{} enabled them to ask questions as they had them, even ``\textit{at night also}'' (ASHA3).
The bot provided instant answers, addressing their doubts immediately. 
ASHAs and ANMs reported that they often used \ashabot{} during their free time, such as while travelling on the bus.
Some ASHAs, especially those with difficulties typing or reading in Hindi, found the audio features for input and output quick and easy to use.
ASHA12 noted a preference for voice messages: ``\textit{It is... difficult to text in Hindi, particularly half letters... There are chances of spelling mistakes.}''
Overall, 26.7\% of ASHAs' questions were in audio format, suggesting the convenience that the audio feature brought to them. 
On the system end, \ashabot{} was robust enough to recognize a few translation and transcription errors while addressing these audio questions, as it received both Hindi and English versions of the question.
For instance, when ASHA5's audio question was incorrectly transcribed as asking about a baby's `\textsf{slow mother-in-law}' ({\dn sAs}, or \textit{saas}) instead of a baby's slow breathing rate ({\dn sA\1s}, or \textit{saans}), \ashabot{} began its response with, `\textsf{I believe there may be a translation error in your question. If you are asking about what to do if the newborn baby's breathing is slow...}'.
On the other hand, 96.8\% of ANMs' responses were text-based. 
They preferred text because it allowed them to carefully revise each part of their response before sending it out, unlike audio messages.
\chimohit{We also observed that both ASHAs and ANMs used Google Keyboard's speech-to-text feature for composing WhatsApp text messages, preferring it over WhatsApp's built-in `voice message' feature due to the latter's less familiar press-and-hold interaction. 
Thus, some text messages sent to \ashabot{} were likely generated from speech inputs.}

ASHAs also appreciated the persistent nature of communication on \ashabot{}, as it allowed them to reread messages when needed.
However, despite the value of being able to access past interactions with the bot anytime, many ASHAs and ANMs had to regularly use the `clear chat' feature on WhatsApp to delete messages across their personal and group chats. 
This was because their phones---provided by \anonymousngo{} over two years ago---often ran out of storage and became slow, especially when logging field data.

\bheading{Overworked ANMs and Doctors}
ASHAs shared that they usually seek answers from ANMs or occasionally from doctors during monthly training sessions.
However, because ANMs and doctors are often busy, ASHAs reported prioritizing their questions and feeling hesitant in asking ``\textit{simple questions}'' (ASHA11). 
ASHAs feared being judged, as they had been scolded in the past for not knowing basic information.
In contrast, \ashabot{} provided a non-judgemental platform, allowing many ASHAs to ask anything without hesitation.
\begin{quote}
    ``\textit{Some questions stay in my mind, I do not ask ANMs or doctors... but with the bot, I can ask anything... (e.g., Once) there was a crowd, it was chaotic, and I mistakenly gave the vaccine (intended) for a 3.5-month-old child to a 2.5 month old and vice versa. I asked \ashabot{}... It said that }`\textsf{You can give the missed vaccine next month... so the babies can get vaccinated within 6 months.}'\textit{ I was relaxed... Otherwise I would have visited those children's homes again and again to make sure that they were okay.}'' -- ASHA10
\end{quote}

With multiple responsibilities and limited time, ANMs and doctors often provide ``\textit{short}'' and ``\textit{cut to cut}'' (ASHA10) answers to ASHAs.
In contrast, \ashabot{} offered detailed, nuanced responses, equipping ASHAs to make informed decisions.
ASHAs felt that they could ask \ashabot{} as many follow-up questions as needed without any concerns for time constraints or expecting impatience, like they had to experience sometimes with ANMs and doctors.
This freedom allowed them to explore topics in depth and revisit information until they fully understood it.
\begin{quote}
    ``\textit{I asked about haemoglobin deficiency... The bot said }`\textsf{If you eat this vegetable, this will happen, eat daal [lentils], eat this, eat that.}'\textit{ It's a detailed explanation, so I understand it well. But ma'am [ANM] only tells me the main point: `}The haemoglobin needs to be at this level... take the medicines, done.'\textit{ The (bot) is better, because the patient might not even need to take medicines.}'' -- ASHA3
\end{quote}
\chimohit{
Older ASHAs found \ashabot{} especially useful for staying up-to-date on new information on vaccinations, medicines, and contraceptive methods.
The bot helped them avoid the embarrassment they sometimes felt when asking questions to ANMs, who were often younger than them.
Reflecting on their own initial struggles, these older ASHAs also acknowledged \ashabot{}'s importance for newer ASHAs, who often lack experience and confidence, and feel hesitant to approach ANMs and doctors.
}

We note that despite ASHAs' shift to using \ashabot{}, it was not seen as a replacement for the feedback and support provided by ANMs, nor did it undermine the ANMs' perceived authority.
Instead, ANMs recognized \ashabot{} as a time-saving tool that could potentially reduce their workload by offering ASHAs the ability to independently resolve their questions.
Many ASHAs noted that their ANMs encouraged them to use the bot not only to address their immediate doubts, but also to expand their knowledge beyond the basics.

\bheading{Asking Sensitive Questions}
\label{findings:rq2:privacy}
ASHAs were often hesitant to discuss sensitive topics, such as women's sexual health, with male doctors and, at times, even with ANMs.
\ashabot{} provided a private channel to address these queries, which many ASHAs found beneficial.
\begin{quote}
    ``\textit{The bot is better... I cannot openly ask the Doctor sir about white discharge, about itching and burning...}'' -- ASHA3
\end{quote}

However, a few ASHAs refrained from asking the bot questions they considered too sensitive.
When asked why, they expressed concerns that \anonymousngo{} might be monitoring their queries, extrapolating from previous experiences where \anonymousngo{} was involved in monitoring their use of other data collection applications. 
Misunderstandings about how \ashabot{} operated exacerbated this issue. 
A few ASHAs mistakenly believed that a human controlled the system, making them hesitant to ask such questions.
For example, ASHA3 typed and then deleted a question about menstruation, assuming that the male field worker who introduced her to \ashabot{} was the one responding.
ASHA17 believed a woman acted as \ashabot{} due to the female voice in the audio messages she received from  \ashabot{}.
Furthermore, because ASHAs and ANMs typically use WhatsApp groups for communication with each other and with \anonymousngo{}, some assumed \ashabot{} could identify them and relay their identity along with the questions to ANMs.
For instance, ASHA16 hesitated to ask about domestic violence, fearing her questions could be traced back to her.
\begin{quote}
    ``\textit{Even though I want to ask, I cannot, because it will take a toll on the woman's family... I was scared that if I ask the bot, you will have all the details that (this) ASHA has asked certain questions... It's in the records... If that gets leaked, then I will be asked more questions.}'' -- ASHA16
\end{quote}

\subsubsection{Barriers to ANMs' Usage}
\label{findings:rq2:anmsnouse}

Most ANMs felt that there were no incentives, financial or otherwise, for them to use \ashabot{}, making it just another online task that added to their effort and screen time.
Some ANMs felt overwhelmed by the number of questions asked to them.
They also felt that some of those questions were too basic and increased their workload unnecessarily.
\begin{quote}
    ``\textit{Sometimes there are so many silly questions about pregnancy and vaccinations... ASHAs have so many years of experience, they should already know these things. Why do they ask such questions?}'' -- ANM4
\end{quote}
This reinforces the ASHAs' concern that they hesitate to ask questions because they fear judgment from their supervisors. 

A fear of accountability also influenced ANMs' use of \ashabot{}.
Some ANMs were reluctant to admit their lack of knowledge of certain answers on record, fearing negative consequences. 
As ANM8 expressed, ``\textit{Everyone will know that I don't know the answer... What if it is a simple question? They will judge me.}''
Consequently, most ANMs chose to ignore the questions they could not answer rather than respond with `\textsf{I don't know}' on the bot.
 
ANMs also struggled to answer questions that were ambiguous or not clearly phrased. 
This was sometimes due to a lack of context, such as when ASHAs asked about medicine dosage without specifying whether the patient was a mother or a newborn (ANM5 and ANM9). 
At other times, the lack of clarity was due to transcription and translation errors. 
38.3\% of the questions sent to ANMs were audio messages, as regional dialects and background noise sometimes hindered the bot's ability to directly answer those.
For instance, when the vaccine `BCG' (Bacillus Calmette-Guérin, a tuberculosis prevention vaccine  used in India~\cite{okafor2024bcg}) was mistakenly transcribed as `BCC', ANM6 chose to ignore the question, assuming it referred to an unfamiliar vaccine.
ASHAs' original audio messages, sent along with the transcribed text, occasionally helped clarify these issues.
When `{\dn vjn}' [\textit{vajan}, meaning weight] was transcribed as `{\dn Bjn}' [\textit{bhajan}, meaning devotional song],
ANMs answered the question correctly by listening to the corresponding audio.
However, the effectiveness of audio was limited, as
ANMs were often reluctant to listen to the audio messages, especially in public spaces where playing it could be ``\textit{awkward}'' (ANM8).
Such scenarios where questions were incomplete or incorrectly formatted often required back-and-forth message exchanges between ASHAs and ANMs, which was not supported by \ashabot{}.
\begin{quote}
    ``\textit{There was a question, }`\textsf{For pregnant women, how much time is safe for walking?}' \textit{It depends on the pregnancy. If it is a high risk case, bed rest is advised... totally depends on the prescription of doctors. I didn't answer... If I answer anything wrong, the chances of risk would rise.}'' -- ANM8
\end{quote}

As a result, ANMs preferred answering ASHAs' questions in person rather than using the bot.
In-person interactions allowed ANMs to better understand ASHAs' queries and address follow-up questions, enabling them to provide more detailed answers, compared to the ``\textit{brief or one-word}'' (ANM8) responses given via the bot.
A minority even felt the bot reduced their individual liability for providing accurate and complete answers, as it aggregated responses from multiple ANMs.


Finally, when responding to questions, ANMs found \ashabot{}'s default reply, `\textsf{Thank you for your answer \includegraphics[height=0.65\baselineskip]{Figure/emoji_pray.pdf}}' unsatisfactory, as it did not offer meaningful feedback. 
They complained that the bot thanked them regardless of the content, even when they responded with `\textsf{I do not know}' or provided incorrect information. 
ANMs wanted to know whether their answers were correct, and if not, what the correct answer was.
They desired constructive feedback to improve their knowledge.
When we clarified that \ashabot{} sought their input because it lacked the correct answers, ANMs suggested that they should at least be able to view the responses provided by other ANMs.


\subsection{Bot's Role in Addressing the Information Needs of ASHAs}
\label{findings:rq3}

\subsubsection{Learning}
\label{findings:rq3:learning}

\ashabot{} contributed to ASHAs' learning experience by addressing their immediate knowledge needs and helping them confirm familiar information. 
Older ASHAs reported that it allowed them to revisit and reinforce knowledge gained from previous training sessions.
\chimohit{ASHAs also noted that the bot was particularly useful for confirming specific numerical details. 
For instance, the Centchroman contraceptive tablet~\cite{jha2024chhayatabletcentchroman} has a complex dosage schedule that changes weekly, making it difficult to recall.} 

Beyond routine work, ASHAs used the bot to explore advanced topics, equipping themselves with knowledge that might be useful in future.
For example, ASHA2 consulted \ashabot{} on `\textsf{home-based care}' for babies with low birth weight.
She recalled feeling unprepared in the past when families could not afford hospital treatment for their child and wanted to be ready ``\textit{just in case}'' similar situations arise again.
\chimohit{ASHAs also valued the bot's ability to address questions beyond public health, such as about government schemes or financial planning. 
With \ashabot{}, they could provide immediate information to care recipients, without redirecting them to other sources of information, such as banks or post offices.}
Even outside of work, ASHAs sometimes sought information on \ashabot{} prompted by their personal experiences.
For instance, after witnessing a heart attack at a local temple, ASHA3 used the bot to gain clarity on the incident: ``\textit{What could have happened? He died so suddenly... it was on my mind. So, when I came back... I asked questions about heart attack.}''

The Related Questions feature proved beneficial for ASHA workers, enabling them to ask follow-up questions
and deepen their understanding.
Usage logs revealed that 59.1\% of the questions asked by ASHAs were Related Questions, which they found challenging to formulate on their own.
This feature also introduced ASHAs to medical terminology they were otherwise unfamiliar with, such as ``\textit{toxoplasmosis}'' (ASHA13).

The resulting increase in knowledge improved ASHAs' practical readiness for monthly training sessions, making them more active and confident participants. 
As ASHA16 noted:
\begin{quote}
    ``\textit{[In these training sessions] We get asked a lot of questions, related to HBNC, geriatrics, etc., and I could answer them... It (The bot) increased my confidence.}'' -- ASHA16
\end{quote}

\bheading{ASHAs' Strategies to Handle \ashabot{}'s `\textsf{I don't know}' Responses}
\ashabot{} responded with `\textsf{I don't know}' to 16.4\% of ASHAs' questions. 
In 92.8\% of such cases, ASHAs did not receive a subsequent consensus answer from ANMs too. 
They then employed alternative strategies based on their existing knowledge and the urgency of the question.
For critical questions, ASHAs consulted other sources, such as ASHA handbooks, online resources, ANMs, or doctors.
\begin{quote}
    ``\textit{I asked the bot about accidentally giving too much iron syrup to infants... It is hard to give the correct dosage when they move their hands... The bot could not answer, so I asked the CHO [doctor], who reassured me that... an accidental extra drop or two wouldn't cause harm.}'' -- ASHA16
\end{quote}
When ASHAs had partial knowledge, they relied on their existing information to continue their tasks, often retrying the bot later. 
For instance, ASHA4 thrice asked for a list of hospitals offering free treatment under a specific government scheme.
When \ashabot{} could not provide an answer, she directed her care recipients to the two hospitals she knew.
Finally, for questions prompted by curiosity rather than immediate need, ASHAs typically accepted `\textsf{I don't know}' responses without further action. 
For example, ASHA9, a yoga practitioner, inquired about appropriate levels of exercise for pregnant women. 
When \ashabot{} could not provide an answer, she did not attempt to seek the information elsewhere.

\bheading{ANMs' Learning}
\ashabot{} also contributed to learning for ANMs.
ANM4 even regarded \ashabot{} as a productive use of her smartphone, remarking that it ``\textit{reduces my YouTube time.}''
Although ANMs were generally familiar with the answers to most of the questions raised on \ashabot{}, when ASHAs initiated detailed follow-ups and advanced medical queries, often through Related Questions, ANMs were encouraged to reflect deeply, expand their knowledge and ``\textit{stay updated}'' (ANM8).
These complex questions often led ANMs to consult senior experts or conduct online research, which they then shared. 
For instance, when ASHA6 asked `\textsf{What precautions should be taken while donating eyes?}', ANM1, who viewed \ashabot{} as a ``\textit{challenge}'' to test both her knowledge and resourcefulness, consulted a doctor at the local hospital.
She conveyed her learnings via \ashabot{}: `\textsf{The donation must be completed within four hours of death}'.

Despite this proactive approach, the new knowledge gained by ANMs was not always communicated back to the bot.
For example, when ANM3 researched `\textsf{What are the symptoms of ectopic pregnancy?}' on YouTube, she understood the answer but did not add it to \ashabot{} because the information was too lengthy to summarize, and the video format did not allow for easy referencing, requiring her to draft an answer from scratch, for which she had little time. 

\bheading{Collective Sensemaking}
\ashabot{} facilitated collective understanding among both ASHAs and ANMs. 
ASHAs reported that their peers, who were not directly participating in the study, would relay their questions through them to the bot.
For instance, ASHA3, while discussing how to protect newborns from heatwaves with a group of ASHAs, used \ashabot{} to inquire about the topic and played the audio response aloud:
\begin{quote}
    ``\textit{I just raised the question, but all 16-17 ASHAs around me learned from the answer.}'' -- ASHA3
\end{quote}
Among the ANMs, ANM5 and ANM9, who were neighbours, often collaborated by discussing and answering \ashabot{} questions using each other's phones. 
For instance, when ANM5 received a question about the purpose of folic acid, they collectively researched the topic on Google and then provided a response.
Through this exercise, ANM9 learned that folic acid helps increase red blood cell counts, which she then added as a response on ANM5's phone.
Overall, most ANMs indicated that they did not hesitate to ask for help when responding to \ashabot{} questions.

\subsubsection{Trust, Expertise, and Authoritativeness}
\label{findings:rq3:trust}
ASHAs considered the answers provided by \ashabot{} to be trustworthy and dependable.
This confidence developed as ASHAs initially tested the bot with questions they already knew the answers to, and compared the bot's responses with their own knowledge.
Sometimes when uncertain, they would verify \ashabot{}'s answers by consulting ANMs, doctors, ASHA handbooks, or online sources, and consistently found the bot's answers to be ``\textit{correct}'' (ASHA2) and ``\textit{complete}'' (ASHA6).

As a result, ASHAs gradually reduced the frequency with which they double-checked the bot's answers and began to rely on it directly.
For example, ASHA13 used \ashabot{} while visiting the home of an anaemic pregnant woman.
When the woman asked for specific dietary recommendations, the ASHA asked the bot, which suggested foods such as `\textsf{\textit{daal} [lentils], rice, egg, and fish}'.
She promptly shared this information with the woman.


ASHAs and ANMs also appreciated the quality and precision of the information that \ashabot{} shared.
They contrasted the bot with other tools they used to find answers such as ``\textit{Google}'' or ``\textit{YouTube}''.
As ANM1 noted, ``\textit{If I Google a question, it does not give me a short answer... It gives the whole Ramayana }[a lengthy Indian epic]\textit{ of the topic, which is not useful to me}.''
ANM9 observed that ASHAs often chose to click on the simplest options during Internet searches, and consumed information with little regard for its reliability. 
In contrast, she felt that \ashabot{} would provide more trustworthy and appropriate responses. 

Interestingly, \ashabot{}'s `\textsf{I don't know}' responses also contributed to building trust.
ASHAs appreciated that the bot either provided accurate information or admitted when it did not know the answer.
However, \ashabot{}'s transparency was a double-edged sword.
`\textsf{I don't know}' responses to ASHAs' audio messages included the text transcription of their corresponding audios. 
Reading this, the ASHAs recognized that the bot sometimes misunderstood them.
For some ASHAs, this lowered their confidence in the bot's ability to meet their information needs.
A few even avoided asking critical questions via audio, fearing misinterpretation.
For example, ASHA6, who predominantly used the audio modality, preferred asking her ANM directly about which vaccination to use rather than consulting \ashabot{}. 
She feared that a misunderstanding on behalf of the bot could lead to her administering the wrong vaccination.

ASHAs noted that if \ashabot{} and their ANM provided different answers, they were more inclined to rely on the bot. 
\ashabot{}'s onboarding messages emphasised its partnership with \anonymousngo{} and the Ministry of Health and Family Welfare, Government of Rajasthan. 
This reinforced ASHAs' trust, as they believed that ``\textit{whatever the government says is correct.}'' (ASHA17).
Additionally, ASHAs perceived ANMs as having limited knowledge and only slightly more formal education than themselves. 
ASHAs had previously encountered situations where ANMs provided incorrect information, leading to this reduced credibility for ANMs compared to \ashabot{}. 
An ASHA recounted: 
\begin{quote}
    ``\textit{Albendazole is generally given to pregnant women when they are in the 6th month... but that ANM gave it to one woman when she was only 3 months pregnant... I raised this issue but she said, `}It is okay, nothing will happen.\textit{' I reported this to [doctor] sir during a sector meeting.}'' -- ASHA16
\end{quote}

\chihighlight{
In practice, there were two reported instances, both mentioned by ASHA1---the most active ASHA among our participants---in which \ashabot{}'s answers were inconsistent with the ASHAs' existing knowledge.
\chiaddagain{First, the ASHA believed that iron tablets should be taken at night after food. 
When the bot instead recommended taking them in the morning on an empty stomach, she was unconvinced. 
To verify, she asked her ANM, who supported her original understanding.}
Second, when asked about the responsibilities of an ASHA worker, \ashabot{} mentioned `\textsf{... assisting victims of violence in accessing medical care, counselling, and legal aid.}'
ASHA1 disagreed, believing her role was limited to health issues, \chiaddagain{and did not pursue the topic further with the bot or her ANM}.
In both cases, ASHA1 relied on the existing (though incorrect) human knowledge over \ashabot{}'s factually correct answers.
}

\bheading{Authoritativeness}
Barring ASHA1, all other ASHAs reported no inconsistencies in \ashabot{}'s answers.
In fact, when a doctor provided unsatisfactory guidance to ASHA10, she turned to the bot for verification, treating it as a supplemental authority.
Similarly, \ashabot{} influenced the decisions of care recipients by providing clear and reliable information. 
For example, ASHA3 used the bot to address concerns from two women in her village who were hesitant to receive the Antara injectable contraceptive.
The women were worried that this injection would cause permanent infertility, as they had heard from others about experiencing irregular menstrual cycles after receiving it.
\ashabot{}'s clarification enabled ASHA3 to effectively convince them.
    \begin{quote}
        ``\textit{Because most women don't take Antara... and they [the ones who have taken Antara] ask us about their [irregular] periods... How to make them understand? I asked about this yesterday to the bot. It told me, }`\textsf{Irregularities are just the side effects of the injection... it is temporary and does not cause infertility.}' \textit{In the evening I explained to the two women about this, even showed them the bot's answer... and this morning both of them came to get the injection.}'' -- ASHA3
    \end{quote}
\section{Discussion}
Our study demonstrates the potential of an LLM-powered, ANMs-in-the-loop chatbot to meet the information needs of ASHA workers. 
Through a thorough examination of the factors that shaped the participation of ANMs and ASHAs with \ashabot{}, we now discuss opportunities and challenges in designing LLM-powered chatbots for low-wage, low-literate frontline workers like ASHAs in high-stakes settings. 

\subsection{Accounting for Technodeterminism}
Our findings show that ASHAs used the bot for its accurate, complete, and detailed responses, which helped address their queries, support care decisions, and persuade care recipients.
They were not skeptical about \ashabot{}'s responses; instead, they expressed trust and gratitude, viewing it as an authority.
Prior work shows that community health workers often lack familiarity with AI-driven tools and tend to rely heavily on them~\cite{okolo2021chwperceptionofAIinmhealth}.
Even outside of healthcare, individuals with low digital literacy often over-rely on new technologies~\cite{bp-nimisha} like conversational agents~\cite{farmchat_imwut18}. 
This trust and reliance on AI are particularly pronounced in non-Western settings like ours~\cite{liu_understanding_2024,agarwal_ai_2024}. 
\chiadd{However, such over-reliance can cause serious harm in high-stakes frontline health contexts, particularly since LLMs are prone to generating misleading information~\cite{gould2024checkllm} and culturally inappropriate content~\cite{agarwal_ai_2024}.} 
\chihighlight{
While researchers have proposed developer-centric tools like design playbooks~\cite{hong2021AIplaybook} and documentation protocols~\cite{bender2018datastatements,hutchinson2021accountabilityinmldatasets} to enhance AI safety, the effectiveness of these tools remains underexplored~\cite{berman2024responsibleaitools}, and AI safety continues to be a complex, evolving challenge, particularly in high-stakes, non-Western contexts where formal benchmarks are often absent.
}

\chiadd{
To promote a realistic user understanding of AI capabilities and limitations, it is essential to incorporate AI competency training into user onboarding to ensure safe engagement with AI systems~\cite{kapania2022aiauthority,okolo2021chwperceptionofAIinmhealth,ramjee2024cataractbot}.
Similar to our approach, such training should emphasize the potential for errors in LLM-generated responses and their ability to produce synthetic data. 
It should provide specific examples of possible hallucinations and encourage critical assessment of AI outputs, and prompt users to question AI authority and verify critical information with experts when needed.}
This aligns with broader human-AI design guidelines, emphasizing the need to ``\textit{help the user understand how often the AI system may make mistakes}''~\cite{guidelines-humanai-chi19}. 

\chiadd{Our findings, however, revealed that even with training, some ASHAs tended to overrely on AI-generated outputs, often overlooking the need to verify critical information. 
Addressing this issue requires strategies that help novice users better understand AI's limitations---an area that still poses open research challenges.
Approaches like adding accuracy indicators, making data sources more transparent, or designing prompts that push users to question and validate AI responses need further exploration and validation.}


Our findings also revealed that \ashabot{} was not perceived as undermining the authority of ASHAs in their interactions with care recipients, nor did it diminish the role of ANMs in supporting ASHAs.
Instead, participants regarded \ashabot{} as a resource to improve ASHAs' effectiveness and reduce ANMs' workload.
As LLM-powered tools become more integrated into everyday services, it is critical that these technologies are viewed as complements, not replacements, for knowledge workers.
For instance in the \ashabot{} deployment, the support ANMs offer to ASHAs extends well beyond answering questions---it includes mentorship, emotional support, and guidance in navigating complex healthcare situations in messy, resource-constrained environments.
The designers and developers of AI technologies must recognize that technology cannot replace human judgment and care, a concern emphasized in prior HCI4D scholarship~\cite{toyama2017geekheresy, yadav2019breastfeedingchatbot}.
Moving forward, LLM-powered bots should be framed as tools that assist end users, working to ``\textit{amplify human intent and capacity}''~\cite{okolo2021chwperceptionofAIinmhealth, toyama2017geekheresy} rather than as solutions that can independently address complex societal problems.
By positioning these bots as supplemental, fallible tools, we can maintain trust in human elements while still benefiting from the efficiencies that AI offers.

\subsection{Multimodality and Personalization}
 Prior work has demonstrated that speech can be an effective medium for users with low levels of digital literacy such as ASHAs and ANMs~\cite{farmchat_imwut18,ramjee2024cataractbot,findlater2009semiliterateaudiotext,vashistha_voice_2023}.
While \ashabot{} allowed ANMs and ASHAs to interact using audio and textual modality, our deployment revealed that they faced several challenges in using audio inputs and outputs. To begin with, errors in translation and transcription significantly impacted usability and usefulness, with some ASHAs being so concerned about potential misinterpretation by the system that they even avoided asking critical questions via audio.
Furthermore, the inability to review and edit audio messages before sending them affected ANMs' use of the audio features.
Also, ASHAs and ANMs hesitated to speak (or play audio responses) out loud and had concerns about being overheard. 
These privacy concerns were evident during interviews, where ASHAs and ANMs avoided sensitive topics or spoke in hushed tones, mirroring patterns noted by~\citet{ismail2018solidarity}.

While ASHAs and ANMs struggled with the audio, there might be unique opportunities to promote other forms of multi-modal interactions with the bot. 
For example, research has shown that visual aids can be highly effective for users with limited literacy~\cite{medhi2006textfreeforlowliterate}, offering an intuitive way to interact.
Recent advancements in multimodal LLMs~\cite{openai2024gpt4o} can make it feasible for bots to process and respond to image-based inputs, which ASHAs can use to submit photos and videos of growth charts, breastfeeding positions, medicine dosages, or even the data they collect. 
Once \ashabot{} has access to such information, it could analyze images to provide relevant answers or additional visual resources.
Furthermore, by querying the data that knowledge workers collect, a multi-modal bot could provide them with timely, relevant, and personalized information in an accessible manner. 
For instance, the bot could integrate with users' work calendars---which, in the case of ASHAs, include key dates like pregnancy checkups, childbirth, vaccinations, and training meetings---and send reminders at critical moments.

\chiadd{\chihighlightagain{However, as the bot gains visibility within the frontline healthcare ecosystem, it could amplify invisible work~\cite{ming_invisible_2022} as well as tensions between stakeholders.
For instance, the government may seek to leverage the bot as a tool for monitoring ASHAs' learning and performance, in order to decide incentives and inform training programs.
This objective could directly conflict with ASHAs' desire to ask the bot sensitive questions which they feel less comfortable asking to their supervisors.
If ASHAs perceive that their queries or mistakes are under scrutiny, it could deter them from fully engaging with--and benefiting from--the system.
This tension between users' privacy needs and authorities' monitoring goals has been noted in prior work in multi-stakeholder healthcare ~\cite{jo2024healthchatbotsmultistakeholder} and educational settings~\cite{varanasi_investigating_2021}.}
}
Hence while incorporating personalization features, 
it is essential to ensure that users have both awareness and control over how much of their collected data and past conversations are shared with an LLM-powered system \chiadd{\chihighlightagain{and other stakeholders}}. 
Balancing personalization with ethical considerations in AI for health remains an ongoing discussion~\cite{maeckelberghe2023ethicalaihealth,okolo2021chwperceptionofAIinmhealth,yadav2019breastfeedingchatbot}, which becomes even more complex when working with low-literate frontline health workers~\cite{ismail2018solidarity,ismail2019empowermentonmargins}.

\subsection{Value Alignment}
While facilitating collaboration between LLM-powered bots and healthcare workers, it is crucial to ensure that the underlying models have a deep understanding of the sociocultural context in which they are used. 
For instance, we found that some ASHAs used \ashabot{} not only to ask health-related questions but also to seek guidance on social issues, such as domestic violence and child marriage. 
In such critical scenarios, it is essential that the values embedded in the LLMs align with those of the ASHAs and ANMs, ensuring that the responses are culturally appropriate.
\chiadd{
Without this value alignment, there is a risk of LLMs imposing Western values and norms~\cite{agarwal_ai_2024}, potentially causing harm, especially when addressing complex social issues such as patriarchy, as highlighted by \citet{sultana_design_2018}.}


The use of technology for frontline workers has predominantly focused on tracking, monitoring, and evaluating them for rewards and compensation~\cite{karunasena2021datacollectiondiligence,whidden2018feedbackdashboard,Henry2016supervisionwhatsappkenya,derenzi2012smstoimproveperformance}.
In our study, some ASHAs mistakenly assumed that their questions were being monitored, causing hesitation in asking sensitive or basic questions.
Likewise, ANMs were concerned about being held accountable for providing incorrect answers and were reluctant to admit knowledge gaps.
A few ASHAs asked as many questions as possible, while ANMs provided numerous answers, both assuming their interactions would be used to evaluate them.
While we clarified these misconceptions during the interviews, the ways users engaged with the bot was influenced by the values \chiadd{and apprehensions} around surveillance and tracking, instead of using the bot as a learning tool.

These findings highlight the importance of aligning AI systems with the values of frontline health workers to ensure that the technology fosters trust, rather than fear or anxiety. 
To achieve this alignment, future deployments should clearly communicate the purpose of the technology from the outset, ensuring that frontline workers understand it is designed to support learning, not to track or monitor performance. 
This proactive approach will help mitigate misconceptions and create a more supportive environment for the growth and development of frontline workers.

\subsection{Governance and Accountability on Scale}

Our findings show that ASHAs derived significant value from using the \ashabot{}, highlighting its potential for broader deployment. 
\chiadd{\chihighlightagain{
However, there are several open questions before scaling up such LLM-powered chatbots. For example: \textit{ Who governs what data should be fed into the bot and keep accountability in check? Whose voices and values should be prioritized in a multi-stakeholder environment? Who is accountable for potential mistakes and value misalignment? How can we prevent additional burdens on already overstretched ASHAs and ANMs?}}}

\citet{xiao2023aichatbotexpertsourcing} emphasizes that domain expertise is essential for identifying and curating reliable and clear content for information portals. In our pilot study, medical doctors from our research team managed and updated the bot's knowledge base, the field team ensured that the bot’s responses matched the linguistic styles of ASHAs and ANMs, and the development team ensured that the bot prioritized local values instead of enforcing Western norms. 
While this approach ensured that the bot was deployed in a safe and controlled environment, new governance models, fairness frameworks, and safety approaches are needed to scale such bots at a population level, especially in a country like India with over 1 million frontline health workers \chiadd{operating in diverse cultural and socioeconomic contexts.}

\chiadd{\chihighlightagain{ If not done with care, scaling could introduce tensions between government priorities and the information needs of ASHAs and their care recipients, particularly on sensitive topics such as family planning. 
As noted in prior work, decisions about what constitutes `appropriate' LLM responses can inadvertently privilege certain perspectives over others~\cite{jang2024llmsforautisticworkers}.
For instance, government-issued healthcare guidelines might emphasize the benefits and ease of contraceptive methods while downplaying potential side effects~\cite{hartmann2024indiaspopulationprogram,geampana2016pregnancymoredangerousthanpill}, creating a risk of biased or incomplete information. Moreover, 
ASHAs' need for detailed, regionally specific information could conflict with the government's preference for consistent, standardized messaging.}}
Hence, for large-scale deployment, it is crucial to center the voices and values of community health workers 
and use participatory approaches to design and develop emerging AI technologies \emph{effectively}. 
In doing so, it is essential to be mindful of whose values go into the design and development of AI technologies, ensure that the designers and builders account for power imbalances (e.g., between ASHAs and their supervisors) and avoid tokenism, where input is sought but not genuinely considered in decision-making~\cite{delgado_participatory_2023}. 

Additionally, preventing catastrophic failures, such as hallucinations, requires leveraging the “expert-in-the-loop” approach. 
In our deployment, this did not work as well as expected---many ANMs hesitated to provide answers to \ashabot{}’s queries due to time constraints, lack of knowledge, or concerns about accountability. 
This raises concerns about the burden placed on an already overstretched workforce \chiadd{\chihighlightagain{and underscores the need to align ASHAs’ needs with their supervisors’ capacities realistically.}}
While previous research has compensated experts financially for their contributions to healthcare chatbots~\cite{karusala2023chatbasedinfoservice}, more work is needed to explore the sustainability of this approach and examine alternative strategies to keep experts motivated, such as peer comparison, which has proven effective in frontline health contexts~\cite{derenzi2017voiceandwebfeedback}. 
\chiadd{\chihighlightagain{
Finally, it is crucial to limit the demands on experts, for example, by setting daily limits on how much they can contribute to be respectful of their existing workload.  
Reducing unnecessary expert involvement is equally important. 
For instance, to address referrals caused by translation and transcription errors, we recommend offering users the flexibility to rephrase and resubmit unclear questions before forwarding them to experts. 
It is also important to 
incorporate expert input into the knowledge base, enabling the bot to answer more questions independently over time. 
As seen in prior work on expert-in-the-loop LLM chatbots~\cite{sachdeva2024learningslargescaledeploymentllmpowered}, this approach is likely to progressively reduce supervisors' bot-related workload.
}}

\subsection{Limitations}
Our exploratory work has some limitations.
First, we employed a small sample size and interview-based methods to assess the feasibility of \ashabot{} and understand the individual experiences of ASHAs and ANMs.
While our findings suggest that \ashabot{} effectively met the information needs of these ASHAs, participant bias~\cite{dell2012participantbias} may have influenced the results.
Second, the study is limited to a specific geographic area within a single state in India, which may restrict the generalizability of the findings.
The nuances of user experiences may vary if the bot is scaled across community workers in other parts of India or in other Global South contexts. 
Future research should involve a larger, geographically diverse group of community workers and examine their usage over an extended period to develop more robust quantitative insights, such as understanding how ASHAs' questions evolve over time and the subsequent impact on ANMs' workload and attitudes.
\section{Conclusion}
ASHAs are essential to India's healthcare system, yet the infrastructure to improve their limited medical knowledge and skill is inadequate. 
To tackle this, we used a LLM-powered experts-in-the-loop open-source framework, to design and develop \ashabot{}.
\ashabot{} is a WhatsApp-based chatbot that uses a doctor-curated knowledge base to provide ASHAs with instant answers to their questions.
When uncertain, the bot consults multiple ANMs and generates a consensus response, which is also used to enhance the knowledge base. 
After extensive pilot testing to ensure our system's accuracy and appropriateness, we conducted a \chidelete{in-the-wild }\chiadd{field} study involving 20 ASHAs and 15 ANMs.
Our interviews with them, and analysis of their interaction log data, highlighted the effectiveness of \ashabot{} in addressing ASHAs' information needs.
ASHAs valued the bot as a convenient and private channel for them to ask rudimentary or sensitive questions without fear of judgement.
They generally trusted the bot, especially as it admitted when it did not know an answer.
Eventually, \ashabot{} played an authoritative role, influencing the decisions of both ASHAs and their care recipients.
On the ANM side, they appreciated the chance to review and expand their knowledge by providing answers on \ashabot{}, but felt overwhelmed by the workload and accountability, suggesting they may need incentives to contribute regularly.
Overall, we find that an LLM-powered chatbot can significantly improve access to information for those with limited literacy and technological exposure in healthcare roles.
Finally, we caution against over-reliance and emphasize the need to position these resources for community health workers as a supplement, and not a replacement, for their supervisors.

\begin{acks}
We express our sincere gratitude to the participants for their time and insights. 
This work is supported by Microsoft and Khushi Baby.
Aditya Vashistha's contributions to this project are supported by the President's Council of Cornell Women.
\end{acks}

\bibliographystyle{ACM-Reference-Format}
\bibliography{main}

\appendix
\section{Appendix}

\subsection{ANM Consensus Prompt}
\label{appendix:prompts:consensus}

\subsubsection{System Prompt}
    \#\#\#Task Description:\\
    A question ("q") asked by an Accredited Social Health Activist (ASHA) has been answered by multiple Auxiliary Nurse Midwives (ANMs) ("anm\_answers"). Your task is to synthesise facts and clarifications in anm\_answers into a simple and comprehensive answer ("consensus\_answer"), by first identifying any conflicting information in anm\_answers and providing a count ("anm\_votes"), and second generating a precise explanation ("consensus\_explanation").\\    
    \#\#\#Steps:\\
    1. Read q and anm\_answers carefully. \\
    2. Identify information within anm\_answers relevant to q. Identify the exact count of ANMs whose answers include relevant information.\\
    3a. If the count is less than 3: (A) Provide an empty string for anm\_votes. (B) Provide only "Consensus not reached." as consensus\_answer.\\
    3b. Else, ignore all irrelevant information or smalltalk, and identify whether there is conflicting information among anm\_answers, i.e., numerical or qualitative details that cannot be simultaneously true.\\
    3b.i. If there are no conflicts: (A) Provide an empty string for anm\_votes. (B) Provide consensus\_explanation. (C) Provide consensus\_answer.\\
    3b.ii. Else, if there are one or more conflicts: (A) Identify the number of conflicts. (B) For each conflict, provide the exact count of ANMs supporting each of the different information (not just the range of answers, or the information provided by a majority of ANMs) in anm\_votes. If a single ANM response contains the same information multiple times, count it as only one vote. (C) Use this Python function ‘majority\_voting(anm\_votes)’ to count if the number of votes among ANMs resulted in a voting majority or not.\\
    def majority\_voting(anm\_votes):
        results = {}
        for conflict, votes in anm\_votes.items():
            \# Convert the vote counts from strings to integers
            votes = {key: int(value) for key, value in votes.items()}
            \# Find the maximum number of votes received
            max\_votes = max(votes.values())
            \# Check how many pieces of information have the maximum vote count
            max\_vote\_keys = [key for key, value in votes.items() if value == max\_votes]
            if len(max\_vote\_keys) == 1:
                \# If one piece of information has the highest count, it's the majority
                results[conflict] = "Voting majority"
            else:
                \# If there's a tie, indicate no voting majority
                results[conflict] = "No voting majority"
        return results\\
     3b.ii.1. If there is ‘No voting majority’ (i.e., if there is a tie in votes) for one or more conflicts: (A) Provide consensus\_explanation. (B) Provide only "Consensus not reached." as consensus\_answer.\\
    3b.ii.2. Else, if there is 'Voting majority' for all conflicts: (A) For each conflict, identify the information provided by a majority of ANMs. (B) Provide consensus\_explanation. (C) For each conflict, integrate only that majority information into consensus\_answer. Do not mention minority information or conflicts in consensus\_answer.\\    
    \#\#\#Instructions:\\
    1. Only use information in anm\_answers to generate consensus\_answer. Do not use any other source.\\
    2. Strictly follow the JSON output format in the examples. Do not generate any other (opening or closing) explanations or code.\\
    3. Your output (anm\_votes, consensus\_explanation, and consensus\_answer) must only be in English. The input (q and anm\_answers) can be in Hindi, English, or Hinglish.\\
    4. consensus\_answer must explain information in simple terms without using medical jargon or uncommon words. \\
    5. consensus\_answer must be as short as possible.\\
    6. consensus\_answer must be framed as an answer for ASHA workers, who are not patients themselves. \\
    7. consensus\_explanation must only be 1-2 sentences long.\\
    8. Do not allow the length of the ANM answers to influence your output.\\
    9. Be as objective as possible.\\
    10. Make sure you read and understand these instructions carefully.\\
    11. Keep this document open while reviewing, and refer to it as needed.\\
    12. Think step-by-step.\\    
    \#\#\#Examples:\\    
    \#\#Example 1:\\    
    \#Input:\\
    \{\\
    "q": "{\dn mAlA en kF V\?bl\?V k\4s\? i-t\?mAl kr\?{\qva}{\rs ?\re}}",\\
    "anm\_answers": ["{\dn yhA\2 gB\0 EnroDk goElyA\2 hotF h\4 Ejsm\?{\qva} \rn{21} golF sP\?d r\2g kF ev\2 \rn{7} golF kAl\? r\2g kF hotF h\4 VoVl \rn{28} golF hotF h\4 \3FEwTm bAr \7{f}! krt\? smy mhAvArF k\? pA\2cv\?{\qva} Edn s\? -VAV\0 krnF h\4 aOr EbnA EksF zkAvV k\? \3FEw(y\?k Edn EnDA\0Ert smy pr hF golF l\?nF h\4 agr EksF Edn golF l\?nA \8{B}l jAe to j\4s\? hF yAd aAtA h\4 \7{t}r\2t golF l\?nA h\4 -VAV\0 kF trP s\? golF -VAV\0 krnF h\4 aOr ek p\381wA K(m hot\? hF \8{d}srA -VAV\0 krnA h\4 golF ko b\3CEwo{\qva} s\? \8{d}r rKnA h\4}", "{\dn yh ek \7{s}rE\322wt gB\0 EnroDk sADn h\4{\rs ,\re} EpryX aAn\? k\4 pA\2cv\? Edn s\? \7{f}! krt\? h\4 hr roj ek golF KAnF h}", "haa", "Daily ek goli", "Mala N ke tab.mahila ke period ke 5ve din par lene h, Ansar shi h kiya", "Mala and tablet MC ke paanchvein din se tablet Ke Piche Teer ka Nishan se chalu karni hai Lal tablet MC ke Samay per leni hai", "Per Day 1 tablet"]\\
    \}\\
    \#Output:\\
    \{\\
    "anm\_votes": "",\\
    "consensus\_explanation": "This answer synthesises the unanimous guidance provided by ANMs on the correct usage of Mala N tablets, focusing on starting the cycle, daily intake, and handling missed doses. As the information given by ANMs was qualitatively different but can be simultaneously true, there was no conflicting information and counting votes and identifying the majority was not required.",\\
    "consensus\_answer": "Mala N tablets should be started on the fifth day of the menstrual cycle. The pack contains 28 pills, with 21 white and 7 black pills. One pill should be taken daily at the same time without any interruption. If a pill is missed, take it as soon as you remember. After finishing one pack, start the next pack immediately. Keep the pills out of reach of children."\\
    \}\\    
    \#\#Example 2:\\    
    \#Input:\\
    \{\\
    "q": "HIV walo ko pension milti h kya?",\\
    "anm\_answers": ["No", "Yes", "haa", "{\dn \rn{5} lAK tk kA ilAj srkArF yA \3FEwAiv\?V hA\<E-pVlo m\? En\7{f}Sk EkyA jAy\?gA}", "Han ji"]\\
    \}\\    
    \#Output:\\
    \{\\
    "anm\_votes": {"Pension": {"No": "1", "Yes: "3" }},
    "consensus\_explanation": "The majority of responses indicate that individuals living with HIV do receive pensions. The information about free treatment provides additional support and is included as part of the comprehensive answer.",\\
    "consensus\_answer": "People living with HIV are eligible for pensions and can receive free treatment up to Rs 5 lakh in government or private hospitals."\\
    \}\\    
    \#\#Example 3:\\    
    \#Input:\\
    \{\\
    "q": "How much money will we get for nasbandi and when we will get it?",\\
    "anm\_answers": ["Under the Janani Suraksha Yojana, ASHA workers will receive Rs 600 for transport and other costs when they help women at the hospital. In 2 months you will get.", "{\dn \rn{1} mhFn\? m\?{\qva} sAt sO zpy\? sr}\includegraphics[height=0.65\baselineskip]{Figure/emoji_pray.pdf}\includegraphics[height=0.65\baselineskip]{Figure/emoji_pray.pdf}\includegraphics[height=0.65\baselineskip]{Figure/emoji_pray.pdf}\includegraphics[height=0.65\baselineskip]{Figure/emoji_pray.pdf}", "haan 600 in 2 months", "ASHA will get 600 rs for ur travel and other costs when you stay with women at the hospital.", "600 In 1 month"]\\
    \}\\
    \#Output:\\
    \{\\
    "anm\_votes": {"Money for sterilisation": {"Rs 700": "1","Rs 600": "4"}, "When they would get it": {"1 month": "2", "2 months": "2" }}, \\
    "consensus\_explanation": "There was confusion among ANM responses regarding the amount and timing of payments for sterilisation support. Although a majority of ANMs agreed that ASH workers will receive support of Rs 600 for their travel and other costs, however variation in the timeframe for receiving the payment (1 month vs. 2 months) prevented a unified answer.",\\
    "consensus\_answer": "Consensus not reached."\\
    \}
\subsubsection{Query Prompt}
XX

\subsection{Medical Answer Evaluation Prompt}
\label{appendix:prompts:evaluation}

    \#\#\#Task Description:\\
    1. Evaluate the quality of a student answer ("Ans") on a scale of 1-3, for each of the below six distinct "Evaluation Metrics", by comparing it with a teacher’s reference answer ("Ref") that scores "3" on all metrics.\\
    2. While evaluating, begin with a short and precise one-sentence "Explanation" of your rating for each metric, strictly based on the definitions given below. Using the Explanation, output your final “Rating” for each metric. \\    
    \#\#\#Instructions:\\
    1. Strictly follow the JSON output format in the Examples below. Do not generate any other (opening or closing) explanations or code.\\
    2. For each metric that involves comparison, only compare Ans with Ref. Do not use any other source.\\
    3. Do not allow the length of the answers to influence your evaluation.\\
    4. Be as objective as possible.\\
    5. Make sure you read and understand these instructions carefully. \\
    6. Keep this document open while reviewing, and refer to it as needed.\\
    7. Think step-by-step.\\    
    \#\#\#Evaluation Metrics:\\
    1. Accuracy: Correctness of information in Ans compared to Ref. Note: if Ans is missing information that is present in Ref, or if Ans repeats information, do not consider this while evaluating Accuracy. Also note: this scale uses "NA" instead of "2".\\
    1a. "1": Ans contains some information directly contradicted by Ref.\\
    1b. "NA": Ans includes information not present in Ref, making their accuracy indeterminate.\\
    1c. "3": All information in Ans is directly supported by Ref.\\    
    2. Subset: Degree to which the set of information in Ans is contained within the set of information in Ref. Note: if Ans is missing some information that is present in Ref, or if Ans repeats information, do not consider this while evaluating Subset.\\
    2a. "1": None of the information in Ans is present in Ref. Ans and Ref are disjoint sets of information. \\
    2b. "2": Ans includes some information not present in Ref. Either Ans and Ref are intersecting sets of information, or the information in Ref is a subset of the information in Ans.\\
    2c. "3": All information in Ans is present in Ref. Ans is a subset of Ref.\\    
    3. Completeness: Degree to which the set of information in Ref is contained within the set of information in Ans. Note: if Ref is missing some information that is present in Ans, or if Ans repeats information, do not consider this while evaluating Completeness.\\
    3a. "1": None of the information in Ref is present in Ans. Ref and Ans are disjoint sets of information.\\ 
    3b. "2": Ref includes some information not present in Ans. Either Ref and Ans are intersecting sets of information, or the information in Ans is a subset of the information in Ref.\\
    3c. "3":  All information in Ref is present in Ans. Ref is a subset of Ans.\\    
    4. Conciseness: Degree of repetition in Ans. Note: This metric does not require comparison with Ref.\\
    4a. "1": All information in Ans is mentioned twice or more.\\
    4b. “2”: Some information in Ans is mentioned twice or more.\\
    4c. "3": All information in Ans is mentioned only once. \\    
    5. Clarity: Degree to which Ans explains (medical) information in simple terms without using jargon or uncommon words, making it understandable to those without a medical background. Note: This metric does not require comparison with Ref.\\
    5a. "1": Ans relies entirely on medical jargon or uncommon words without providing explanations in layman’s terms.\\
    5b. "2": Ans primarily uses simple language but includes some medical jargon or uncommon words without providing explanations in layman’s terms.\\
    5c. "3": Ans primarily uses simple language. Medical jargon or uncommon words are either not included at all, or accompanied with explanations in layman’s terms.\\    
    6. Structure: Degree of logical arrangement of information in Ans compared to Ref. Note: If Ans is missing some information that is present in Ref, or if Ref is missing some information that is present in Ans, do not consider this while evaluating Structure. Only consider the set of information that is present in both Ref and Ans. \\
    6a. "1": Ans does not align with the logical arrangement of any information present in Ref. This is also applicable when Ans and Ref are disjoint sets of information. \\
    6b. "2": Ans aligns with the logical arrangement of some information present in Ref. Some information in Ans are logically arranged in the same way as in Ref, but others are not. \\
    6c. "3": Ans aligns with the logical arrangement of all information in Ref.\\    
    \#\#\#Examples:\\    
    \#\#Example 1:\\    
    \#Input:\\    
    \{\\
    "Ref": "The eye drops prescribed are to be applied only in the eye that has undergone surgery, in the lower fornix, i.e., the fold between the back of the lower eyelid and the eyeball. However, if you have any doubts, it's always best to consult with your doctor. If you already have been advised eye drops for the other eye, it is best to continue the same as per advice.",\\
    "Ans": "The eye drops prescribed are typically to be applied in the eye that has undergone surgery, in the lower fornix. However, it's always best to follow the specific instructions given by your doctor. If you're unsure, please consult with your doctor."\\
    \}\\
    \#Output:\\
    \{\\
    "Accuracy": \{\\
    "Explanation": "All information in Ans (the use of eye drops in the operated eye) is directly supported by Ref; although Ans is missing some information that is present in Ref (whether prescribed drops should continue in the other eye), omissions in Ans do not affect the evaluation of Accuracy.",\\
    "Rating": "3"\\
    \},\\
    "Subset": \{\\
    "Explanation": "All information in Ans is present in Ref; Ans is a subset of Ref; although Ans  is missing some information that is present in Ref (about existing prescriptions), omissions in Ans do not affect the evaluation of Subset.",\\
    "Rating": "3"\\
    \},\\
    "Completeness": \{\\
    "Explanation": "Ref includes some intersecting information with Ans (the use of eye drops in the operated eye) but Ref also includes some information not present in Ans (continuation of existing treatment for the other eye and an explanation of the term ‘fornix’).",\\
    "Rating": "2"\\
    \},\\
    "Conciseness": \{\\
    "Explanation": "All information in Ans (the use of eye drops in the operated eye and the suggestion to consult with a doctor) is mentioned only once.",\\
    "Rating": "3"\\
    \},\\
    "Clarity": \{\\
    "Explanation": "Ans primarily uses simple language, but includes a medical jargon ‘fornix’ without providing any explanation.",\\
    "Rating": "2"\\
    \},\\
    "Structure": \{\\
    "Explanation": "Ans aligns with the logical arrangement of all information in Ref that is present in both Ref and Ans (instruction followed by the suggestion to consult with a doctor).",\\
    "Rating": "3"\\
    \}\\
    \}\\    
    \#\#Example 2:\\    
    \#Input:\\
    \{\\
    "Ref": "We are in the process of enabling online scheduling. In the meanwhile please schedule your appointment by calling the number 9739270477 or writing in to pc.blr@sankaraeye.com.",\\
    "Ans": "You can schedule your appointment by contacting the hospital at sankaraeye.com or writing in to info@sankaraeye.com. You could also call during working hours on +9197399270477."\\
    \}\\    
    \#Output:\\
    \{\\
    "Accuracy": \{\\
     "Explanation": "The phone number in Ans is directly contradicted by the phone number in Ref, as Ans includes an extra digit '9' which might be a typing mistake. \\
     "Rating": "1"\\
    \},\\
    "Subset": \{\\
    "Explanation": "Ans includes some intersecting information with Ref (calling and email as contact methods); but Ans also includes some information not present in Ref (website, different email, wrong phone number, and the advice to call during working hours).",\\
    "Rating": "2"\\
    \},\\
    "Completeness": \{\\
    "Explanation": "Ref includes some intersecting information with Ans (calling and email as contact methods); but Ref also includes some information not present in Ans (correct phone number,email ID, and the fact that online scheduling is yet to be enabled.)",\\
    "Rating": "2"\\
    \},\\
    "Conciseness": \{\\
    "Explanation": "All information in Ans (website, email ID, and phone number) is mentioned only once.",\\
    "Rating": "3"\\
    \},\\
    "Clarity": \{\\
    "Explanation": "Ans primarily uses simple language to explain the contact methods with no medical jargon or uncommon words",\\
    "Rating": "3"\\
    \},\\
    "Structure": \{\\
    "Explanation": "Ans mentions the email option first, and the phone call option second, while Ref is ordered conversely.",\\
    "Rating": "1"\\
    \}\\
    \}\\
    \#\#Example 3:\\    
    \#Input:\\
    \{\\
    "Ref": "For a cataract surgery under topical or local anaesthesia, you can have a light breakfast at your usual time. It's ideal to ensure around a 45-minute gap between your breakfast and being wheeled into the operation theatre, especially if your surgery is scheduled early. However, if the surgery is under general anaesthesia, an empty stomach is recommended.\\ This would be advised by the anaesthesia team.",
    "Ans": "If the surgery is planned under general anaesthesia, the patient should check with the attending staff or anaesthetist on when the last meal could be had. It is always best to consult with them when under general anaesthesia. It's generally recommended to have a light meal before coming for the surgery."\\
    \}\\    
    \#Output:\\
    \{\\
    "Accuracy": \{\\
    "Explanation": "Ans directly contradicts Ref by saying that it's recommended to have a light meal before surgery under general anaesthesia, while Ref states that an empty stomach is recommended for the same.",\\
     "Rating": "1"\\
    \},\\
    "Subset": \{\\
    "Explanation": "None of the information in Ans about food intake prior to general anaesthesia is present in Ref.",\\
    "Rating": "1"\\
    \},\\
    "Completeness": \{\\
    "Explanation": "None of the information in Ref about food intake recommendations for either anaesthesia prior to cataract surgery is present in Ans.",\\
    "Rating": "1"\\
    \},\\
    "Conciseness": \{\\
    "Explanation": "Ans twice mentions the advice to consult with medical staff when under general anaesthesia.",\\
    "Rating": "2"\\
    \},\\
    "Clarity": \{\\
    "Explanation": "Ans primarily uses simple language while explaining food intake recommendations and does not include any medical jargon or uncommon words.",
    "Rating": "3"\\
    \},\\
    "Structure": \{\\
    "Explanation": "Ans and Ref are disjoint sets of information and therefore Ans does not follow the logical structure of any information present in Ref; Ans focuses only on general anaesthesia while Ref makes different recommendations based on the type of anaesthesia.",\\
    "Rating": "1"\\
    \}\\

\end{document}